\documentclass[a4paper,11pt]{article}
\usepackage{jheppub} % for details on the use of the package, please see the JINST-author-manual

\usepackage{jheppub}
\usepackage{amsmath}
\usepackage{amssymb}
\usepackage{appendix}
\usepackage[table]{xcolor}
\usepackage{multirow}
\usepackage{rotating}
\usepackage{diagbox}
\usepackage{slashed}
\usepackage{float}
\usepackage{cancel}
\usepackage{tabularx}
\usepackage{tikz}
\usepackage{bbold}
\usepackage{tikz-feynman}
\usepackage{arydshln}

\usetikzlibrary{calc,tikzmark,fit,shapes.geometric,matrix,decorations.markings,arrows.meta,decorations.pathmorphing,patterns,positioning,snakes}
\usepackage[capitalize]{cleveref}
\usepackage[normalem]{ulem}

\newcommand\supsetsim{\mathrel{\substack{\textstyle\supset\\[-0.2ex]\textstyle\sim}}}

\title{\boldmath A common origin of the Higgs boson and the flavor hierarchies}

\author{Javier M. Lizana}
\affiliation{Instituto de F\'isica Te\'orica UAM/CSIC, Nicolas Cabrera 13-15, Madrid 28049, Spain}

\emailAdd{jmlizana@ift.csic.es}

\abstract{We present a model that extends the electroweak gauge symmetry of the Standard Model in a non-universal way to $SU(2)_{L}^{\prime}\times U(1)_X \times SU(2)_{L}^{q_3}\times SU(2)_R^{\ell_3}$. This symmetry is spontaneously broken to $SU(2)_L\times U(1)_Y$ near the TeV scale by a condensate of a new composite sector. Charging appropriately the fermionic degrees of freedom of the composite sector, anomaly cancellation enforces the Standard Model fermions to be charged in such a way that the extended gauge interactions respect a $U(2)_q\times U(2)_e\times U(3)_u\times U(3)_d\times U(3)_\ell$ accidental flavor symmetry. 
In addition, from the same symmetry breaking, a composite Higgs boson emerges as a pseudo-Nambu-Goldstone boson of the strong dynamics of the new sector.
Due to the extended gauge and the specific flavor symmetry, leading Yukawa couplings between Higgs and fermions can only be written for the third generation and higher dimension operators generate suppressed light-family Yukawa couplings. Furthermore, CKM mixing angles between third and light families are naturally suppressed while the PMNS ones, anarchic.
The model thus provides a unified origin for the Higgs boson and the flavor hierarchies between third and light families.
}
\preprint{IFT-UAM/CSIC-24-185}
\makeatletter
\gdef\@fpheader{}
\makeatother

\begin{document}
\maketitle
\flushbottom

\section{Introduction}
\label{sec:intro}

The Higgs sector of the Standard Model (SM) harbors many of the open puzzles of particle physics. The fact that its mass is not protected from radiative corrections of heavier new physics (NP) is a clear hint for NP close to the electroweak (EW) scale. The look for the stabilization of the Higgs mass has been one of the main driving forces for model building and a well-exploited guiding principle to suggest ultraviolet (UV) completions of the SM. Among the different proposals, composite Higgs models are one of the most appealing solutions~\cite{Kaplan:1983fs,Panico:2015jxa}. According to them, the Higgs is a resonance of a new composite sector. Scale invariance, only broken by dimensional transmutation, protects the scale of confinement of the new sector from UV physics.

The reason behind the hierarchies of the Yukawa couplings of the Higgs boson to the SM fermions, which translate into the hierarchies of masses and mixing matrices, constitutes another open puzzle, the flavor puzzle~\cite{Altmannshofer:2024jyv}. 
Quarks and charged leptons are arranged into three generations or families with increasing masses, expanding almost 6 orders of magnitude, with a difference between the different generations of about 2 orders of magnitude. Also, the Cabibbo-Kobayashi-Maskawa (CKM) matrix is quite aligned to the identity, with particularly suppressed mixing elements between light and third families. 
Such a particular pattern seems to ask for an underlying dynamical explanation.
Neutrinos however exhibit a very different pattern. Not only their masses are 6 orders of magnitude below the electron mass, but the Pontecorvo–Maki–Nakagawa–Sakata (PMNS) matrix has an anarchic structure, with O(1) mixing elements. 
Part of the this puzzle is precisely understanding why CKM and PMNS matrices are so different.

These flavor hierarchies result in very suppressed Yukawa couplings to the light generation, endowing the SM with approximate, but almost exact, $U(2)$ flavor symmetries suppressing flavor changing neutral currents (FCNC) and CP violating processes. For this reason, observables sensitive to this kind of processes, which are in very good agreement to the SM, impose very strong constraints to NP beyond the SM, specially if it resides close to the EW scale, as it is necessary to address the Higgs mass stabilization. Although these bounds can be avoided if NP is universal, searches from LEP and LHC push the scale of universal NP to $\sim 10$ TeV, making more difficult to find a satisfactory explanation for the lightness of the Higgs.

An intelligent approach to overcome these difficulties is to consider NP models that implement the same approximate $U(2)$ flavor symmetries than the SM, and thus flavor observables are protected. Note that the breaking of universality between third and light families is experimentally allowed at the TeV scale~\cite{Allwicher:2023shc}.
These $U(2)$ symmetries could be originated from some far UV dynamics via for example, horizontal gauge groups~\cite{Froggatt:1978nt,Pomarol:1995xc,King:2003rf,Buras:2011wi,Greljo:2023bix,Greljo:2024zrj}, but they could also emerge directly as accidental symmetries at the TeV scale. This opens the very appealing possibility of addressing (partially) the flavor puzzle at the scale of stabilization of the Higgs, at least between third and light families. The hierarchy between the first and second families however has probably to be postponed to the PeV scale or higher as suggested by meson-mixing experimental limits. The emergence of the flavor hierarchies could therefore be a consequence of the existence of several NP scales, a scenario which has been proposed and studied in several contexts~\cite{Berezhiani:1983hm,Berezhiani:1992pj,Barbieri:1994cx,Panico:2016ull,Bordone:2017bld,Allwicher:2020esa,Fuentes-Martin:2020pww,Fuentes-Martin:2022xnb}.

It is been pointed out that imposing an approximate $U(2)$ flavor symmetry on the light left-handed (LH) quark doublets, $U(2)_q$, results in suppressed masses for first and second family quarks, and suppressed mixing angles between these and the third family~\cite{Greljo:2023bix,Davighi:2023xqn,Capdevila:2024gki}. However, in the lepton sector, the analogous $U(2)_{\ell}$ on LH lepton doublets to suppress masses of light leptons imposes selection rules on the Weinberg operator and has to be extended with more structure to explain the anarchy of the PMNS matrix. It is more minimal to impose the $U(2)$ symmetry on the RH charged leptons, $U(2)_e$~\cite{Antusch:2023shi}.

Flavor deconstruction refers to UV completions where the SM gauge group is split into multiple factors under which SM fermions transform non-universally~\cite{Li:1981nk,Muller:1996dj,Malkawi:1996fs,Cheng:2001vd,Shu:2006mm,Chiang:2009kb,Hsieh:2010zr,Craig:2011yk,Chivukula:2013kw,Bordone:2017bld,Greljo:2018tuh,Crosas:2022quq,Allwicher:2023aql,Davighi:2023iks,FernandezNavarro:2023rhv,Davighi:2023evx,FernandezNavarro:2023hrf,Davighi:2023xqn,Barbieri:2023qpf,Capdevila:2024gki,FernandezNavarro:2024hnv}. These UV completions can be used to implement accidentally the discussed flavor symmetries. 
A common feature of all these models is the apparition of new massive vector bosons from the breaking of the non-universal gauge symmetry with very interesting phenomenology.
Models combining flavor deconstruction with a composite Higgs have also been proposed~\cite{Fuentes-Martin:2020bnh,Chung:2023gcm,Covone:2024elw}.
If the extended gauge group is broken to the SM by scalar fields, to avoid gauge anomalies, complete families are typically charged under the different factors of the gauge symmetry, limiting the possible accidental symmetries that can emerge. For instance, achieving $U(2)_q\times U(2)_e$ while the other SM fields remain $U(3)$ is forbidden this way.
A radiatively stable alternative to scalars is triggering the breaking by the condensate of a new composite sector~\cite{Fuentes-Martin:2020bnh}. Furthermore, a specific charge assignment of the fermionic degrees of freedom of the new sector can lead to a split of same-family fermions into different factors of the gauge group~\cite{Fuentes-Martin:2024fpx}, enlarging the number of possibilities of the emerging flavor symmetries.

In this paper we present a model with the accidental flavor symmetry $U(2)_q\times U(2)_e$ emerging naturally at the TeV scale via flavor deconstruction of the EW gauge group with a breaking triggered by the condensate of a composite sector with EW charges. The model is free of gauge anomalies and offers an optimum starting point to address the flavor hierarchies both in the quark and lepton sectors.
Furthermore, not all the composite pseudo-Nambu-Goldstone bosons (pNGB) of this breaking are eaten by the massive vector bosons of the extended gauge, but nine scalars, arranged as one singlet and two $SU(2)_L$ doublets with the same quantum numbers than the Higgs boson, survive as physical pNGBs. One of the doublets can develop an EW vacuum expectation value (VEV), breaking the EW symmetry and becoming the SM Higgs.
Due to the extended gauge, this composite Higgs can acquire leading Yukawa couplings only to third family fields. Higher dimensional operators can generate suppressed light-family Yukawa couplings, in a way that the CKM mixing elements are naturally suppressed, but PMNS ones, remain $O(1)$.
The model thus unifies the emergence of a composite Higgs boson with the breaking of the deconstructed gauge symmetry, which is at the core of the explanation of the flavor hierarchies.

This paper is organized as follows. In \cref{sec:Model} we establish the guiding principles we follow, and use them to build our model. \cref{sec:Higgs} is devoted to analyze how the Higgs boson emerges from our construction and its main properties. In \cref{sec:Pheno} we explore the main phenomenological consequences of our model, find the lower experimental limit of the scale of NP, and discuss on future experiments capable to further test the model. We conclude in \cref{sec:Conclusions}.

\section{Building the model}

\label{sec:Model}

Regarding the quark sector of the SM, imposing the $U(2)_q$ flavor symmetry between the light LH quark doublets allows for non-suppressed Yukawas involving only the third family quark doublet,
\begin{equation}
-\mathcal{L} \supset \sum_{i=1}^3 \left(y^{t}_i \,\bar q_L^3 \tilde H u^i_R + y^{b}_i \,\bar q_L^3  H d^i_R \right).
\end{equation}
These Yukawa couplings break the accidental $U(3)_u\times U(3)_d$ to $U(2)_u\times U(2)_d$, and define the third-family RH quarks: $\sum_i y^t_i u^i_R  = y_t t_R$ and $\sum_i y^b_i d^i_R  = y_b b_R$. The remaining  Yukawa couplings have to be written by inserting spurions of $U(2)_q$, $V_{q,t(b)}$ and $\Delta_{u(d)}$,
\begin{equation}
-\mathcal{L} \supset \sum_{i=1}^2 y_{t}V_{q,t}^i  \,\bar q_L^i \tilde H t_R +\sum_{i=1}^2 y_{b}V_{q,b}^i \,\bar q^i_L  H b_R + 
\sum_{i,j=1}^2 y_{t}\Delta_u^{ij}  \,\bar q_L^i \tilde H u^j_R +\sum_{i,j=1}^2 y_{b}\Delta_d^{ij}  \,\bar q_L^i  H d^j_R .
\end{equation}
If this $U(2)_q$ symmetry is implemented by some UV completion of the SM, these spurions are expected to be suppressed by the new dynamics, suppressing also light-quark masses and CKM mixing elements between third and light families.
However, the large mixing angles in the PMNS matrix disfavor the analogous $U(2)_{\ell}$ in the lepton sector. Although it can explain the hierarchy between the $\tau$ mass and the other charged leptons in a similar way, it imposes selection rules on the Weinberg operator that suppress the PMNS mixing elements:
\begin{equation}
-{\cal L} \supset   \frac{1}{\Lambda_{\nu}} \bigg[
y_{\nu_3}(\bar \ell^3_L \tilde H) (H^{\dagger} \ell^{3c}_L)
+\sum_{i=1,2} y^{\nu}_{3i} (\bar \ell^3_L   \tilde H) (H^{\dagger}\ell^{i c}_L)
+ \sum_{i,j=1,2} y^{\nu}_{ij} (\bar \ell^j_L  \tilde H) (H^{\dagger}\ell^{i c}_L)
\bigg].
\end{equation}
Based on the spurion analysis, one expects $1\sim y_{\nu_3} \gg y^{\nu}_{3i} \gg y^{\nu}_{ij}$, implying $m_{\nu_2}/m_{\nu_3} \ll \theta_{23}\ll 1$. Extra structure has to be added to suppress $y_{\nu_3}$ and $y^{\nu}_{3i}$ to make them of the order of $y^{\nu}_{ij} $ and achieve anarchy in the neutrino sector. 
As pointed out in Ref.~\cite{Antusch:2023shi}, it is more minimal to impose a $U(2)_{e}$ flavor symmetry between the light RH charged leptons, suppressing light charged lepton Yukawas without implying any selection rule in the Weinberg operator, and therefore, any hierarchy in the PMNS matrix.

A possible UV completion capable of imposing this kind of flavor symmetries is flavor deconstruction. Deconstructions of $SU(2)_L$, i.e. extensions of the EW gauge group to $SU(2)_1\times SU(2)_2 \times U(1)_Y$, where light-quark doublets transform under $SU(2)_1$ and the third-family-quark doublet under $SU(2)_2$, impose the $U(2)_q$ symmetry if the Higgs is charged under $SU(2)_2$~\cite{Davighi:2023xqn,Capdevila:2024gki}. 
However, in standard flavor deconstructions using a scalar field to break the UV symmetry group to the SM one, and without extra fermion fields, anomaly cancellation imposes to charge lepton doublets in the same way, leading to a $U(2)_{\ell}$ symmetry. Otherwise, mixed gauge anomalies $U(1)_Y-SU(2)_{Li}-SU(2)_{Li}$ are generated.
Although an anarchic PMNS matrix can still be obtained by further extending hypercharge to $U(1)_R\times U(1)_{B-L}$ and additionally deconstructing $U(1)_{B-L}$~\cite{Greljo:2024ovt},
in this paper we explore a different path. Applying the ideas of anomalous flavor deconstruction of Ref.~\cite{Fuentes-Martin:2024fpx}, it is possible to achieve flavor deconstructions with an accidental $U(2)_q\times U(2)_e$ flavor symmetry.

\subsection{Anomalous flavor deconstruction}
\label{sec:AFD}

The hypercharge gauge group $U(1)_Y$ can be seen as the diagonal subgroup of the $SU(2)_R\times U(1)_{B-L}$ global group.\footnote{Furthermore, $U(1)_{B-L}\times SU(3)_c$ can be embedded into the $SU(4)$ group, realizing the Pati-Salam unification~\cite{Pati:1974yy}.} Similarly to LH fields of the SM that are $SU(2)_L$ doublets, RH fields are then arranged into doublets of $SU(2)_R$, where we need to include RH neutrinos, $\ell^i_R = (\nu^i_R,e^i_R)$. 
We here explore deconstructions~\cite{Hill:2000mu,Arkani-Hamed:2001nha} at the TeV scale of the $SU(2)_L\times SU(2)_R$ part of the group that may endow the model with the $U(2)_q\times U(2)_e\times U(3)_u\times U(3)_d\times U(3)_\ell$ accidental symmetry. Our models then have the approximate global symmetry
\begin{equation}
\mathcal{G}_{2^41}=SU(2)_{L1} \times SU(2)_{R1} \times SU(2)_{L2} \times SU(2)_{R2}\times U(1)_{B-L}.\label{eq:G241}
\end{equation}
Although in a far UV, the full group could be gauge, we here choose to gauge at the TeV scale a smaller subgroup:
\begin{equation}
\mathcal{G}_{2^31}= SU(2)_{L1}\times U(1)_X \times SU(2)_{L2} \times SU(2)_{R2} ,
\end{equation}
where $U(1)_X = [SU(2)_{R1}\times U(1)_{B-L}]_{\rm diag}$ and with gauge couplings $g_{L1}$, $g_{L2}$, $g_{X}$ and $g_{R2}$. 
Alternatively we could gauge the subgroup $U(1)_{R2}$, instead of the full $SU(2)_{R2}$. 
This would lead to a very similar model with some differences we will comment in \cref{foot:U1R0,foot:U1R}. 
We however prefer to gauge maximally $SU(2)_{R2}$ to absorb as many pNGBs as possible and minimize the number of $U(1)$ factors of our gauge group.

Following the ideas of~\cite{Fuentes-Martin:2020bnh,Fuentes-Martin:2024fpx}, 
the breaking of the UV gauge symmetry to the SM one will be triggered by the condensate of a new confining sector. 
This has the advantage of been radiatively stable versus other options like link scalar fields.
To build it, we introduce 4 hyper-quarks $\zeta_{L,R}$ transforming in a vector-like fundamental representation of a new gauge sector $SU(N_{\rm HC})$ with $N_{\rm HC}$ hyper-colors. We arrange the hyper-quarks as doublets of the $SU(2)$ factors of ${\cal G}_{2^41}$: $\zeta_L=(\zeta^{(L)}_L,\zeta^{(R)}_L)$, where $\zeta^{(L)}_L$ is a doublet of $SU(2)_{L1}$ and 
$\zeta^{(R)}_L$ a doublet of $SU(2)_{R1}$, and $\zeta_R=(\zeta^{(L)}_R,\zeta^{(R)}_R)$, where
$\zeta^{(L)}_R$ is a doublet of $SU(2)_{L2}$,
and $\zeta^{(R)}_R$ a doublet of $SU(2)_{R2}$.
The new confining sector actually has a larger global symmetry: $SU(4)_{1}\times SU(4)_{2}\times U(1)_{\rm HC}$, where $SU(4)_{1(2)}$ are complex rotations acting on the LH (RH) chiralities of the hyper-quarks, $U(1)_{\rm HC}$ a vector-like phase transformation\footnote{Axial-like phase transformations are anomalous.}, and $SU(2)_{L1(2)}\times SU(2)_{R1(2)} \subset SU(4)_{1(2)}$.
At the confining scale $\Lambda_{\rm HC}$, hyper-quarks form a condensate $\langle \bar \zeta^a_L \zeta^b_R\rangle \propto \delta^{ab}$ that breaks the symmetry
\begin{equation}
SU(4)_{1}\times SU(4)_{2}\times U(1)_{\rm HC} \to 
SU(4)_{V}\times U(1)_{\rm HC},\label{eq:HSGlobalSymmBr}
\end{equation}
which is translated into the breaking of
\begin{equation}
SU(2)_{L1}\times SU(2)_{L2} \times SU(2)_{R1} \times SU(2)_{R2} \to SU(2)_L\times SU(2)_R,
\end{equation}
and the breaking of the gauge symmetry $\mathcal{G}_{2^31}$ to the EW one, $SU(2)_L\times U(1)_Y$. 

We want now to charge the SM fermions so gauge interactions have a $U(2)_q\times U(2)_e\times U(3)_u\times U(3)_d\times U(3)_\ell$ symmetry.
For this we should:
\begin{enumerate}
\item[(1)] Split light and thrid-family LH quarks under the two $SU(2)_{Li}$ factors.
\item[(2)] Charge all RH quarks under the same $SU(2)_{Ri}$ factor.
\item[(3)] Split light and thrid-family RH leptons under the two $SU(2)_{Ri}$ factors.
\item[(4)] Charge all LH leptons under the same $SU(2)_{Li}$ factor.
\end{enumerate}
Any arrangement satisfying these conditions in general contributes to mixed anomalies between $U(1)_{B-L}$ and the $SU(2)$ factors, which, if not compensated, will be inherited by the gauge group ${\cal G}_{2^31}$.
However, these anomalies can be canceled by charging appropriately the hyper-quarks under $U(1)_{B-L}$, making the theory anomaly-free~\cite{Fuentes-Martin:2024fpx}.
Two possible models can be built realizing the conditions (1)--(4), but we will select one of them by the following extra condition:
\begin{enumerate}
\item[(5)] All light-family SM fields are charged under the same factors $SU(2)_{L1}$ and $SU(2)_{R1}$.
\end{enumerate}
This condition is desirable for phenomenological reasons: in the limit in which the gauge couplings satisfy $g_{L1}\ll g_{L2}$ and $g_{X}\ll g_{R2}$, the extra vector bosons will decouple from light fields. This allows to decrease their masses, and therefore, the breaking scale, being still compatible with experimental constraints.

The matter content of the model consistent with conditions (1)--(5), and their representations under the global group $\mathcal{G}_{2^41}\times SU(3)_c\times SU(N_{\rm HC})$ and under the gauge group $\mathcal{G}_{2^31}\times SU(3)_c\times SU(N_{\rm HC})$ are explicitly given in \cref{tab:fieldcontent}.
%%%%%
\begin{table}[t]
\renewcommand{\arraystretch}{1.2}
\begin{center}
\begin{tabular}{|c||c|c||c|c||c|c||c|}
\hline
Field  & $SU(2)_{L1}$ & $SU(2)_{R1}$ & $SU(2)_{L2}$ &  $SU(2)_{R2}$ & $U(1)_{B-L}$ & $SU(3)_c$ & $SU(N_{\rm HC})$\\
\hline
\hline
$q^3_L$  & $\mathbf{1}$ & $\mathbf{1}$ & $\mathbf{2}$ & $\mathbf{1}$ & $1/6$ & $\mathbf{3}$ & $\mathbf{1}$\\
\hline
$q^{1,2}_L$  & $\mathbf{2}$ & $\mathbf{1}$ & $\mathbf{1}$ & $\mathbf{1}$ & $1/6$ & $\mathbf{3}$ & $\mathbf{1}$\\
\hline
$q^{1,2,3}_R$ & $\mathbf{1}$ & $\mathbf{2}$ & $\mathbf{1}$ & $\mathbf{1}$ & $1/6$ & $\mathbf{3}$ & $\mathbf{1}$  \\
\hline
$\ell^{1,2,3}_L$  & $\mathbf{2}$ & $\mathbf{1}$ & $\mathbf{1}$ & $\mathbf{1}$ & $-1/2$ & $\mathbf{1}$ & $\mathbf{1}$\\
\hline
$\ell^3_R$  & $\mathbf{1}$ & $\mathbf{1}$ & $\mathbf{1}$ & $\mathbf{2}$ & $-1/2$ & $\mathbf{1}$& $\mathbf{1}$ \\
\hline
$\ell^{1,2}_R$ & $\mathbf{1}$ & $\mathbf{2}$ & $\mathbf{1}$ & $\mathbf{1}$ & $-1/2$ & $\mathbf{1}$  & $\mathbf{1}$\\
\hline
\hline
$\zeta_L^{(L)}$ & $\mathbf{2}$ & $\mathbf{1}$ & $\mathbf{1}$ & $\mathbf{1}$ & $1/2N_{\rm HC}$ & $\mathbf{1}$ &  $\square$ \\
\hline
$\zeta^{(R)}_L$  & $\mathbf{1}$ & $\mathbf{2}$ & $\mathbf{1}$  & $\mathbf{1}$ & $1/2N_{\rm HC}$ & $\mathbf{1}$ & $\square$ \\
\hline
$\zeta^{(L)}_R$ & $\mathbf{1}$  & $\mathbf{1}$ & $\mathbf{2}$ & $\mathbf{1}$ & $1/2N_{\rm HC}$ & $\mathbf{1}$ & $\square$  \\
\hline
$\zeta^{(R)}_R$ & $\mathbf{1}$ & $\mathbf{1}$ & $\mathbf{1}$ & $\mathbf{2}$ & $1/2N_{\rm HC}$ & $\mathbf{1}$  & $\square$ \\
\hline
\hline
$N_L$ & $\mathbf{1}$ & $\mathbf{1}$ & $\mathbf{1}$ & $\mathbf{1}$ & $0$ & $\mathbf{1}$ &  $\mathbf{1}$ \\
\hline
\end{tabular}
\end{center}
%%%%
%%%%
\begin{center}
\begin{tabular}{|c|c|c|c|c|c|c|c|}
\hline
Field  & $SU(2)_{L}^{\prime}$ & $U(1)_{X}$ & $SU(2)_{L}^{q_3}$ & $SU(2)_{R}^{\ell_3}$ & $SU(3)_c$ & $SU(N_{\rm HC})$\\
\hline
\hline
$q^3_L$ & $\mathbf{1}$ & $1/6$  & $\mathbf{2}$ &  $\mathbf{1}$ & $\mathbf{3}$  & $\mathbf{1}$\\
\hline
$q^{1,2}_L$  & $\mathbf{2}$ &  $1/6$ & $\mathbf{1}$ & $\mathbf{1}$  & $\mathbf{3}$ & $\mathbf{1}$\\
\hline
$u^{1,2,3}_R$  & $\mathbf{1}$ & $2/3$ & $\mathbf{1}$ & $\mathbf{1}$ & $\mathbf{3}$ & $\mathbf{1}$ \\
\hline
$d^{1,2,3}_R$  & $\mathbf{1}$ & $-1/3$ & $\mathbf{1}$ & $\mathbf{1}$  & $\mathbf{3}$ & $\mathbf{1}$ \\
\hline
$\ell^{1,2,3}_L$ & $\mathbf{2}$ & $-1/2$ & $\mathbf{1}$ & $\mathbf{1}$ & $\mathbf{1}$ & $\mathbf{1}$ \\
\hline
$\ell^{3}_R$ & $\mathbf{1}$ &  $-1/2$ & $\mathbf{1}$ & $\mathbf{2}$ & $\mathbf{1}$ & $\mathbf{1}$ \\
\hline
$\nu^{1,2}_R$ & $\mathbf{1}$ &  $0$ & $\mathbf{1}$ & $\mathbf{1}$ & $\mathbf{1}$ & $\mathbf{1}$ \\
\hline
$e^{1,2}_R$ & $\mathbf{1}$ &  $-1$ & $\mathbf{1}$ & $\mathbf{1}$ & $\mathbf{1}$ & $\mathbf{1}$ \\
\hline
\hline
$\zeta_L^{(L)}$  & $\mathbf{2}$ & $1/2N_{\rm HC}$ & $\mathbf{1}$ & $\mathbf{1}$ & $\mathbf{1}$ & $\square$\\
\hline
$\zeta_{L,1}^{(R)}$  & $\mathbf{1}$ &  $1/2N_{\rm HC}+1/2$ & $\mathbf{1}$ & $\mathbf{1}$  & $\mathbf{1}$ & $\square$ \\
\hline
$\zeta_{L,2}^{(R)}$  & $\mathbf{1}$ &  $1/2N_{\rm HC}-1/2$ & $\mathbf{1}$ & $\mathbf{1}$ & $\mathbf{1}$ & $\square$ \\
\hline
$\zeta_R^{(L)}$  & $\mathbf{1}$ &  $1/2N_{\rm HC}$  & $\mathbf{2}$ & $\mathbf{1}$ & $\mathbf{1}$ & $\square$\\
\hline
$\zeta_{R}^{(R)}$  & $\mathbf{1}$ &  $1/2N_{\rm HC}$ & $\mathbf{1}$ & $\mathbf{2}$ & $\mathbf{1}$ & $\square$\\
\hline
\hline
$N_L$ & $\mathbf{1}$ & $0$ & $\mathbf{1}$ & $\mathbf{1}$ & $\mathbf{1}$ & $\mathbf{1}$ \\
\hline
\end{tabular}
\end{center}
\caption{Matter content of our model. On the top we arrange the matter fields as representations of the global group ${\cal G}_{2^41}\times SU(3)_c\times SU(3)_{\rm HC}$. On the bottom, we arrange them as representations of the subgroup we gauge, ${\cal G}_{2^31}\times SU(3)_c\times SU(3)_{\rm HC}$. We use $\square$ to denote the fundamental representation.}
\label{tab:fieldcontent}
\end{table}
%%%
The model can be understood as follows. All RH and light LH quarks are located on site 1, while third-family LH quarks are located on site 2, implementing a $U(2)_q\times U(3)_u\times U(3)_d$ flavor symmetry in the quark sector. If hyper-quarks weren't charged under $U(1)_{B-L}$, anomaly cancellation would require a similar arrangement for leptons. However, the $U(1)_{B-L}$ hyper-quark charges of $1/2N_{\rm HC}$ force all LH and light RH leptons to be on site 1, and third-family RH leptons on site 2, completing the desired flavor symmetry.
On top of the SM fermion fields and the hyper-quarks, we have to also include a fermion singlet $N_L$ for reasons that we will explain later in \cref{sec:Neutrinos}.
Given the fields that are charged under each group factor, we rename them as $SU(2)_{L2} \equiv  SU(2)_{L}^{q_3}$, $SU(2)_{R2} \equiv SU(2)_R^{\ell_3}$, and $SU(2)_{L1} \equiv SU(2)_{L}^{\prime}$.
All perturbative anomalies cancel. Furthermore, the cancellation of global or non-perturbative anomalies~\cite{Witten:1982fp} 
requires an even number of fermions in the fundamental representation of every $SU(2)$-factor of the gauge group.
The model is then completely anomaly-free if $N_{\rm HC}$ is odd. Interestingly, this condition implies that hyper-baryons of the hyper-sector will always have integer electric charge, $Q=T_L^3+T_R^3+B-L$. Indeed, hyper-baryons have $B-L=1/2$ and $T_L^3+T_R^3$ is half-integer.
Note that also hyper-mesons have integer electric charge: they have $B-L=0$ and integer $T_L^3+T_R^3$.

One question this model raises is the apparent arbitrariness of the $U(1)_{X}$ charges of the hyper-quarks, necessary to make the model anomaly-free. Is there any UV completion embedding the gauge group of our model into a semisimple group to justify these $U(1)$ charges? In a similar manner that in the SM, the hypercharges of quarks can be justified by embeddings of the SM gauge group into simple or semisimple groups in grand unified theories (GUT) such as the Pati-Salam model~\cite{Pati:1974yy}, something similar happens for our model. In \cref{sec:SemisimpleUV} we show this semisimple UV completion, that could occur in a far~UV.

Importantly, it remains to see how the Higgs boson is implemented so the gauge symmetry indeed suppress light-family Yukawa couplings. This aspect will be discussed in the next section.

The alternative model satisfying conditions (1)--(4) but not (5) is also presented and briefly commented in \cref{sec:AppModelII} for completion.

\section{An emerging composite Higgs}
\label{sec:Higgs}

The breaking $SU(4)_1\times SU(4)_2\times U(1)_{\rm HC}\to SU(4)_V\times U(1)_{\rm HC}$ gives 15 Nambu-Goldstone bosons transforming in the ${\bf 15}$ of $SU(4)_V$. Notice that, due to the $B-L$ charges of the hyper-quarks, $U(1)_{B-L}$ actually corresponds to $[U(1)_{B-L}\times U(1)_{\rm HC}]_{\rm diag}$. Under the unbroken global group $SU(2)_L\times SU(2)_R\times U(1)_{B-L}\subset SU(4)_V\times U(1)_{B-L}$, the Nambu-Goldstone bosons transform as
\begin{equation}
{\bf 15}_0 \sim({\bf 3},{\bf 1})_0+({\bf 1},{\bf 3})_0+2\times ({\bf 2},{\bf 2})_0+({\bf 1},{\bf 1})_0,
\end{equation}
i.e. an $SU(2)_L$ triplet $T^a$, an $SU(2)_R$ triplet $\Delta^a$, a singlet $S$ and two $SU(2)_L\times SU(2)_R$ bi-doublets $H_{1,2}$.
Some of them are eaten by new massive gauge bosons. Under the EW gauge group, these massive gauge bosons are a vector triplet ${\cal W} \sim {\bf 3 }_0$, which eats $T^a$, a neutral vector singlet ${\cal B}_0 \sim {\bf 1}_0$, which eats the neutral component of $\Delta^a$, and a charged vector singlet ${\cal B}_1 \sim {\bf 1}_1$, which eats the charged components of $\Delta^{a}$. 
The spectrum of physical pNGB bosons are then a pure singlet $ S\sim {\bf 1}_0$ and two doublets $H_{1,2}\sim {\bf 2}_{1/2}$, which have the same quantum numbers than the Higgs boson. It is therefore natural to wonder if they could trigger the EW symmetry breaking as in a two Higgs doublet model.
Actually, in the context models where the Higgs boson emerges as a pNBG, the symmetry pattern $SU(4)_1\times SU(4)_2\to SU(4)_V$ has been studied in Refs.~\cite{Schmaltz:2010ac,Vecchi:2015fma,Ma:2015gra,Agugliaro:2019wtf}.\footnote{See also Ref.~\cite{Marzocca:2018wcf} for a fundamental composite Higgs with a pattern with larger groups that contain ours.}
In Ref.~\cite{Schmaltz:2010ac}, a little Higgs model is built where the deconstruction of the gauge group is used to implement a collective symmetry breaking to alleviate the tuning. The breaking in that work is only described effectively below the scale of compositeness.
In Refs.~\cite{Ma:2015gra,Agugliaro:2019wtf}, a fundamental composite Higgs with a similar hyper-quark content to our model is constructed.
One main difference of those works with respect to ours is the gauge group: instead of gauging the SM group where $SU(2)_L\times U(1)_Y\subset SU(4)_V\times U(1)_{B-L}$, we gauge ${\cal G}_{2^31}$, with ${\cal G}_{2^31}\not\subset SU(4)_V\times U(1)_{B-L}$. This restricts the leading Yukawa couplings that the model can develop, generating naturally flavor hierarchies.

\subsection{The CCWZ action}
\label{sec:CCWZ}

The low-energy degrees of freedom of the composite sector are the 15 pNGB of the breaking $SU(4)_1\times SU(4)_2\to SU(4)_V$. They can be arranged into the matrix $U$,
\begin{equation}
U=e^{i\,\Pi/f},~~~~~\Pi=
\frac{1}{2}\begin{pmatrix}
T^{a} \sigma_{a}+\frac{S}{\sqrt{2}}\,\mathbb{1}_{2} && -i\,H_{2\times 2} \\
i\, H_{2\times 2}^{\dagger} && \Delta^{a} \sigma_{a}-\frac{S}{\sqrt{2}}\,\mathbb{1}_{2} 
\end{pmatrix},\label{eq:UMatrix}
\end{equation}
where $\sigma_a$ are the Pauli matrices, $f$ is the decay constant of the breaking, $f = \Lambda_{\rm HC}/g_*$, with $g_* =\,4\pi/\sqrt{N_{\rm HC}}$,
\begin{align}
H_{2\times 2} =& (\tilde H_1+i\tilde H_2 ,H_1+iH_2 ),\nonumber\\
T^{a} \sigma_{a} =& \begin{pmatrix}
T^0 && \sqrt{2} T^+ \\
\sqrt{2} T^- && -T^0
\end{pmatrix},~~
\Delta^{a} \sigma_{a} = \begin{pmatrix}
\Delta^0 && \sqrt{2} \Delta^+ \\
\sqrt{2} \Delta^- && -\Delta^0
\end{pmatrix},\label{eq:H2x2}
\end{align}
and $\tilde H^{\alpha} = \epsilon^{\alpha \beta} (H^\dagger)_\beta$. The global group $SU(4)_1\times SU(4)_2$ acts like
$U\to g_1 U g_2^{\dagger}$, where $g_i \in SU(4)_i$. At leading order, the dynamics of the pNGBs is described by the Coleman-Callan-Wess-Zumino (CCWZ) Lagrangian,~\cite{Coleman:1969sm,Callan:1969sn} or chiral effective Lagrangian,
\begin{equation}
\mathcal{L} \supset f^2 \,{\rm Tr} (D_{\mu} U^{\dagger} D^{\mu} U),\label{eq:CCWZ}
\end{equation}
where 
\begin{align}
D_{\mu} U =\partial_{\mu}U &+  \frac{i}{2} g_{L1}(W_1)_{\mu}^{a} 
\begin{pmatrix} \sigma_{a} && 0 \\ 0 && 0   \end{pmatrix}
U
-\frac{i}{2} g_{L2}(W_2)_{\mu}^{a}\, U 
\begin{pmatrix} \sigma_a && 0 \\ 0 && 0   \end{pmatrix}
\nonumber\\
&+ \frac{i}{2} g_{X} (B^0_1)_{\mu} 
\begin{pmatrix} 0 && 0 \\ 0 && \sigma_3   \end{pmatrix}
U - \frac{i}{2} g_{R2} (B_2)^a_{\mu} \,U
\begin{pmatrix} 0 && 0 \\ 0 && \sigma_a   \end{pmatrix}.
\end{align}
The gauge field $(B_2)^a_{\mu}$ is a triplet of $SU(2)_R$, and will be decomposed as
\begin{align}
(B_2)_{\mu}^{a} \sigma_{a} = \begin{pmatrix}
(B^0_2)_{\mu} && \sqrt{2} {\cal B}_{1\mu} \\
\sqrt{2} {\cal B}_{1\mu}^{*} && -(B^0_2)_{\mu}
\end{pmatrix}.
\end{align}
The fields $T^{a}$, $\Delta^0$ and $\Delta^{\pm}$ are eaten by new massive vector bosons, so in the unitary gauge, $T^{a}=\Delta^a=0$. Expanding the Lagrangian, we obtain mass terms for some of the gauge bosons.
We thus define
\begin{align}
\begin{pmatrix}
W^a_{\mu}\\
{\cal W}^a_{\mu}
\end{pmatrix}=
\begin{pmatrix}
\cos \theta_L && \sin \theta_L \\
-\sin \theta_L && \cos \theta_L
\end{pmatrix}
\begin{pmatrix}
(W_1)^a_{\mu}\\
(W_2)^a_{\mu}
\end{pmatrix},~~
\begin{pmatrix}
B_{\mu}\\
{\cal B}_{0\mu}
\end{pmatrix}=
\begin{pmatrix}
\cos \theta_X && \sin \theta_X \\
-\sin \theta_X && \cos \theta_X
\end{pmatrix}
\begin{pmatrix}
(B^0_1)_{\mu}\\
(B^0_2)_{\mu}
\end{pmatrix},
\end{align}
where
\begin{equation}
\sin \theta_{L}=\frac{g_{L1}}{\sqrt{g_{L1}^2+g_{L2}^2}},~~~
\sin \theta_{X}=\frac{g_{X}}{\sqrt{g_{X}^2+g_{R2}^2}}.\label{eq:thetaLX}
\end{equation}
Fields $W^a_{\mu}$ and $B_{\mu}$ stay massless and correspond to the SM gauge bosons with gauge couplings
\begin{equation}
g_L=\frac{g_{L1}\,g_{L2}}{\sqrt{g_{L1}^2+g_{L2}^2}},~~~~~
g_Y=\frac{g_{X}\,g_{R2}}{\sqrt{g_{X}^2+g_{R2}^2}}.
\end{equation}
Fields ${\cal W}$, ${\cal B}_0$ and ${\cal B}_1$ get masses 
\begin{equation}
M_{\cal W}=f\sqrt{g_{L1}^2+g_{L2}^2}\,,~~~~~M_{{\cal B}_0}=f \sqrt{g_{X}^2+g_{R2}^2}\,,~~~~~M_{{\cal B}_1}={f \,g_{R2}},
\end{equation}
and have couplings to the SM fermions
\begin{align}
-\mathcal{L}\supset& \,\frac{1}{2}\mathcal{W}_{\mu}^a \left[g_{L1}^\prime \sum_{\psi_L \in L_1} \bar \psi_L  \sigma_a \gamma^{\mu} \psi_L+
g_{L2}^\prime\,\bar q_L^3 \sigma_a \gamma^{\mu} q_L^3\right]\nonumber\\
&+\mathcal{B}_{0\mu} \left[
g_{X}^\prime \sum_{\psi \in X} Y_{\psi} \bar \psi  \gamma^{\mu} \psi-
\frac{1}{2}(g_{X}^\prime+g_{R2}^\prime) \bar \tau_R \gamma^{\mu} \tau_R
\right],
\end{align}
where $Y_{\psi}$ is the hypercharge of the field $\psi$, $L_1=\{q_L^{1,2},\ell_L^{i} \}$ and $X=\{q_L^{i},u_R^{i},d_R^{i},\ell_L^{i},e_R^{1,2} \}$ ($i=1,2,3$), and the primed couplings are
\begin{equation}
(g_{L1}^{\prime},g_{L2}^{\prime})=g_L(-\tan \theta_L,\cot \theta_L),~~~
(g_{X}^{\prime},g_{R2}^{\prime})=g_Y(-\tan \theta_X,{\cot} \theta_X).
\end{equation}
Furthermore, the RH neutrino $\nu_R^3$ partner of $\tau_R$ under $SU(2)^{\ell_3}_R$ couples to $\mathcal{B}_0$ and $\mathcal{B}_1$,
\begin{equation}
-\mathcal{L}\supset \,\frac{1}{2}(g_{R2}^\prime-g_{X}^\prime)\mathcal{B}_{0\mu}
\bar \nu^3_R \gamma^{\mu} \nu^3_R
+
\frac{g_{R2}}{\sqrt{2}}(\mathcal{B}_{1\mu}
\bar \nu^3_R \gamma^{\mu} \tau_R + {\rm h.c.}).
\end{equation}
The CCWZ action also gives canonically normalized kinetic terms for the pNGBs and interactions between the pNGBs and the massive vector fields $\mathcal{V}$,
\begin{align}
-\mathcal{L}\supset 
&\frac{1}{4}( g^{\prime}_{L1}+g^{\prime}_{L2}) (H_1^{\dagger}\sigma^a i\overleftrightarrow{D}_{\mu} H_1+H_2^{\dagger}\sigma^a i\overleftrightarrow{D}_{\mu} H_2)\mathcal{W}_a^{\mu}\nonumber\\
&\frac{1}{4}( g^{\prime}_{X}+g^{\prime}_{R2}) (H_1^{\dagger}i\overleftrightarrow{D}_{\mu} H_1+H_2^{\dagger}i\overleftrightarrow{D}_{\mu} H_2
)\mathcal{B}_0^{\mu}\nonumber\\
-&\frac{g_{R2}}{2\sqrt{2}}\left[\epsilon_{\alpha\beta}(H_1^{\alpha}iD_{\mu} H_1^{\beta}+H_2^{\alpha}iD_{\mu}  H_2^{\beta} )\,(\mathcal{B}_1^{\mu})^{*}+{\rm h.c.}\right]+O(\mathcal{V}^2),
\end{align}
where here $D$ is the SM covariant derivative. It is interesting that the Higgses couple to the massive vector bosons as the average of the couplings they would have if they were fields located on one of the two sites. Given the opposite sign of the two couplings, this reduces this coupling by more than a factor 2 with respect to other options where the Higgs and the breaking are not unified, improving constraints from EW precision data.
Also, we obtain higher order derivative interactions between the Higgses,
\begin{align}
\mathcal{L}&\supset 
\,\frac{1}{48f^2} 
D_{\mu}|H_1|^2D^{\mu}|H_1|^2
-\frac{1}{12f^2} |H_1|^2 |D_{\mu}H_1|^2 + (H_1 \leftrightarrow H_2)\nonumber\\
&-\frac{1}{48f^2}\left(D_{\mu}(H_1^{\dagger} H_2)D^{\mu}(H_1^{\dagger} H_2)+{\rm h.c.}\right)-\frac{1}{24f^2}D_{\mu}(H_1^{\dagger} H_2)D^{\mu}(H_2^{\dagger} H_1)\nonumber\\
&
-\frac{1}{24f^2}(H_1^{\dagger}H_2+{\rm h.c.})
(D_{\mu}H_1^{\dagger}D^{\mu}H_2+{\rm h.c.})\nonumber\\
&+ \frac{1}{8f^2}\left(H_1^{\dagger}D_{\mu}H_2 + D_{\mu}H_2^{\dagger}H_1\right)
\left(H_2^{\dagger}D^{\mu}H_1 + D^{\mu}H_1^{\dagger}H_2\right)
.\label{eq:H1DH2}
\end{align}
Note that the unbroken global group $SU(4)_V$ contains the custodial symmetry $SU(2)_L\times SU(2)_R$. However, a generic VEV of the Higgses can break it. Indeed, only $\langle H_{2\times 2}\rangle \propto \mathbb{1}$, or equivalently $\langle H_2 \rangle= c\,\langle H_1 \rangle$, with $c \in \mathbb{R}$,  preserves custodial symmetry. In the other cases, the last line of \cref{eq:H1DH2} gives contributions to the $\rho$ parameter~\cite{Mrazek:2011iu,Ma:2015gra}. We will elaborate more on this in \cref{sec:Scalar}.

\subsection{Yukawa couplings}
\label{sec:Yukawa}

To generate the SM Yukawa couplings in composite-Higgs models, the composite sector must interact with the SM fermions.
Two mechanisms are mostly used in the literature: partial compositeness~\cite{Kaplan:1991dc} and via SM fermion bilinears. In the first case, SM fermions $\psi_{\rm SM}$ have linear interactions with fermionic operators $O_{\psi}$ of the composite sector that interpolate fermionic resonances, ${\cal L} \supset \bar \psi_{\rm SM} O_{\psi}$. SM fermions mix with these resonances, inheriting couplings to the Higgs boson. In principle, an advantage of this mechanism is that for certain strongly-coupled sectors, these fermionic operators could have low scaling dimension. Then, terms $\bar \psi_{\rm SM} O_{\psi}$ could be relevant and even for large Yukawa couplings like the top-quark one, the scale at which they are generated could be arbitrary high.
Also, the composite sector needs to have colored fields to generate colored fermionic resonances that can mix with quarks (and in particular, with the top quark).
Our composite sector however is neutral under color, so implementing partial compositeness requires extending this sector, perhaps with scalars charged both under color and hyper-color~\cite{Sannino:2016sfx,Cacciapaglia:2017cdi,Sannino:2017utc,Agugliaro:2019wtf}. 

The alternative to partial compositeness is introducing interactions of a composite operator $O_H$ that interpolates the Higgs boson with bilinears of the SM fermions $\psi_{L,R}$, ${\cal L}\supset\bar \psi_{L} O_{H} \psi_R$. In this case, if ${\rm dim}(\mathcal{O}_H)>1$ (strict), as seems to be necessary to address the hierarchy problem~\cite{Rattazzi:2008pe,Rattazzi:2010yc}, the term $\bar \psi_{L} O_{H} \psi_R$ is necessarily irrelevant, so the scale at which it is generated cannot be decoupled from the composite scale. In our model, bilinears of the hyper-quarks like 
$\bar\zeta_L^{(L)}\zeta_R^{(R)}$ or $\bar\zeta_L^{(R)}\zeta_R^{(L)}$
interpolate $H_{1,2}$,
so $\bar \psi_{L} O_{H} \psi_R$ can be generated with four-fermion operators.
Thus, we need to anyway assume that there is some extended hyper-color sector generating these four-fermion operators. 
We choose this realization, although we will comment in \cref{sec:EHC} on the partial compositeness possibility with no big differences in the results.
Possible four-fermion operators of the type $\bar \psi_{L} O_{H} \psi_R$ with two SM fields and two hyper-quarks allowed by the gauge symmetry are
\begin{align}
\mathcal{L} & \supset \frac{1}{\Lambda_{t}^2}  \sum_{i=1}^3(\bar q^3_{L,\alpha } u_R^i)\left[\lambda_{t,i} \delta^{\alpha \beta} (\bar \zeta_{L,1}^{(R)} \zeta_{R,\beta}^{(L)}) 
+ \lambda^{\prime}_{t,i} \epsilon^{\alpha \beta} (\bar \zeta_{R,\beta}^{(L)} \zeta_{L,2}^{(R)})\right] \nonumber\\ 
&+ 
\frac{1}{\Lambda_{b}^2}\sum_{i=1}^3
(\bar q^3_{L,\alpha } d_R^i)\left[ \lambda_{b,i}\delta^{\alpha \beta} (\bar \zeta_{L,2}^{(R)} \zeta_{R,\beta}^{(L)}) 
+ \lambda^{\prime}_{b,i} \epsilon^{\alpha \beta} 
(\bar \zeta_{R,\beta}^{(L)} \zeta_{L,1}^{(R)})\right]  \nonumber\\
&+ 
\frac{1}{\Lambda_{\tau}^2}\sum_{i=1}^3
(\bar \ell^i_{L,\alpha } \ell^3_{R,\alpha^{\prime}}) \left[ \lambda_{\tau,i}\delta^{\alpha \beta}\delta^{\alpha^{\prime} \beta^{\prime}} (\bar \zeta_{R,\beta^{\prime}}^{(R)} \zeta_{L,\beta}^{(L)}) 
+\lambda^{\prime}_{\tau,i} \epsilon^{\alpha \beta} \epsilon^{\alpha^{\prime}\beta^{\prime} } 
(\bar \zeta_{L,\beta}^{(L)} \zeta_{R,\beta^{\prime}}^{(R)})
\right] 
+{\rm h.c.}\label{eq:ModelI4ferYuk}
\end{align}
where $\alpha^{(\prime)}$, $\beta^{(\prime)}$ are fundamental $SU(2)$ indices, $\alpha^{(\prime)},\beta^{(\prime)}=1,2$. \cref{tab:Illustration1} illustrates why these are the only operators allowed by the gauge symmetry.
%%%%%
\begin{table}[t]
\renewcommand{\arraystretch}{1.2}
\begin{center}
\begin{tabular}{|c||c|c|}
\hline
 & Site 1 ($\zeta_L$) & Site 2 ($\zeta_R$)  \\
\hline
\hline
$SU(2)_L$  & $q_L^{1,2},~\ell_L^{1,2,3}$ & $q_L^3$  \\
\hline
$SU(2)_R$ & $q_R^{1,2,3},~\ell_R^{1,2}$ & $\ell_R^3$ \\
\hline
\end{tabular}
\end{center}
\caption{Arrangement of the SM fields on the sites 1 and 2 (corresponding to the factors that charge the hyper-quarks $\zeta_L$ and $\zeta_R$ respectively). Since the operator interpolating the Higgs is $\mathcal{O}_H \sim \bar \zeta_L \zeta_R$, it can only couple fields in the diagonals of the table due to the deconstructed gauge symmetry.}
\label{tab:Illustration1}
\end{table}
%%%%%
We assume $\lambda_{\psi,i}$ and $\lambda^{\prime}_{\psi,i}$
are linearly dependent in flavor space, $\lambda^{\prime}_{\psi,i}\propto \lambda_{\psi,i}$, something very natural if their UV origin is a single mediator. Then these terms break $U(3)_{u,d,\ell}$ to $U(2)_{u,d,\ell}$.\footnote{If they are not linearly dependent, $U(2)_{u,d}$ will be broken, and light flavor observables will push the scale of the model to the PeV range.}
We can rotate our basis so the third-family RH quarks and the tauonic lepton doublet are those fields perpendicular to the $U(2)$ flavors,
\begin{equation}
\lambda_t^{(\prime)} t_R=\lambda^{(\prime)} _{t,i}u_R^i,~~~
\lambda_b^{(\prime)}b_R = \lambda^{(\prime)} _{b,i}d_R^i,~~~
\lambda_{\tau}^{(\prime)} \ell^{3}_L=\lambda^{(\prime)} _{\tau,i}\ell_L^i,
\end{equation}
where $(\lambda_f^{(\prime)})^2=\sum_{i=1}^3 (\lambda_{f,i}^{(\prime)})^2$.
Four-fermion operators of \cref{eq:ModelI4ferYuk} induce new terms in the chiral effective Lagrangian, coupling pNGBs to SM fermions. We can estimate them by spurion and dimensional analysis. We use the transformation properties under the global group acting on the hyper-quarks, $SU(4)_1\times SU(4)_2$, of the Wilson coefficients $\lambda \sim ({\bf 4},\bar {\bf 4})$ and $\lambda^{\prime} \sim (\bar{\bf 4}, {\bf 4})$  to construct invariant terms with the pNGB matrix $U$. We then insert the correct power of $\Lambda_{\rm HC}$ and $g_*$ to adjust the dimension of masses and couplings~\cite{Giudice:2007fh,Panico:2015jxa}:
\begin{align}
&\mathcal{L} \supset
\, \frac{c\,\Lambda_{\rm HC}^3}{g_*^2\Lambda_t^2}\,(\bar q^3_{L,\alpha } t_R)\left({\lambda_{t}}\delta^{\alpha\beta}U^{\dagger}_{\beta3}+ \lambda^{\prime}_{t}
\epsilon^{\alpha \beta} U_{4\beta}
\right)
+\frac{c\,\Lambda_{\rm HC}^3}{g_*^2\Lambda_b^2}\,(\bar q^3_{L,\alpha } b_R)\left( \lambda_{b} \delta^{\alpha \beta}U^{\dagger}_{\beta 4}
+ \lambda^{\prime}_{b} \epsilon^{\alpha \beta} 
U_{3 \beta } \right)\nonumber\\
+&
\frac{c\,\Lambda_{\rm HC}^3}{g_*^2\Lambda_{\tau}^2}\,(\bar \ell^{3}_{L,\alpha } \tau_R) 
\left( \lambda_{\tau} \delta^{\alpha \beta}U_{\beta 4}
- \lambda^{\prime}_{\tau} \epsilon^{\alpha \beta} 
U^{\dagger}_{3\beta } \right)+
\frac{c\,\Lambda_{\rm HC}^3}{g_*^2\Lambda_{\tau}^2}\,(\bar \ell^{3}_{L,\alpha } \nu^3_R) 
\left( \lambda_{\tau} \delta^{\alpha \beta}U_{\beta 3}
+ \lambda^{\prime}_{\tau} \epsilon^{\alpha \beta} 
U^{\dagger}_{4\beta } \right)
+{\rm h.c.}\label{eq:Yuk}
\end{align}
Here $c$ is a parameter given by the strong dynamics that we expect to be $O(1)$.
Expanding $U$ in terms of pNGB, we obtain
\begin{align}
\mathcal{L} \supset
&\, \frac{c\,\Lambda_{\rm HC}^2}{\Lambda_t^2}\,(\bar q^3_{L,\alpha } t_R)\left[ 
-\frac{\lambda_{t}+\lambda_t^{\prime}}{2g_*}\tilde H_1 - i\frac{\lambda_{t}-\lambda_t^{\prime}}{2g_*}\tilde H_2 
+O(f^{-1})
\right]
\nonumber\\
+&
\frac{c\,\Lambda_{\rm HC}^2}{\Lambda_b^2}\,(\bar q^3_{L,\alpha } b_R)
\left[ 
-\frac{\lambda_{b}-\lambda_b^{\prime}}{2g_*} H_1 - i\frac{\lambda_{b}+\lambda_b^{\prime}}{2g_*} H_2+O(f^{-1})
\right]
\nonumber\\
+&
\frac{c\,\Lambda_{\rm HC}^2}{\Lambda_{\tau}^2}\,(\bar \ell^{3}_{L,\alpha } \tau_R) 
\left[ 
\frac{\lambda_{\tau}+\lambda_{\tau}^{\prime}}{2g_*}H_1 +  i\frac{\lambda_{\tau}-\lambda_{\tau}^{\prime}}{2g_*} H_2 +O(f^{-1})
\right]\nonumber\\
+&
\frac{c\,\Lambda_{\rm HC}^2}{\Lambda_{\tau}^2}\,(\bar \ell^{3}_{L,\alpha } \nu^3_R) 
\left[ 
\frac{\lambda_{\tau}+\lambda_{\tau}^{\prime}}{2g_*}\tilde H_1 +  i\frac{\lambda_{\tau}-\lambda_{\tau}^{\prime}}{2g_*} \tilde H_2 +O(f^{-1})
\right]
+{\rm h.c.}\label{eq:YukawasH3}
\end{align}
We then expect $\Lambda_{t,b,\tau} \gtrsim \Lambda_{\rm HC}$ and $\lambda_t^{(\prime)}\sim g_* $. 
We can see that, specially for the top Yukawa, we are on the edge of validity of the effective field theory (EFT). We should probably give the UV description of the extended hyper-color sector to properly describe its infrared (IR) effects. Still, we can use this description as a proxy to understand how the flavor hierarchies arise.
The last line of \cref{eq:YukawasH3} gives a $\tau$ Yukawa to the RH neutrino $\nu_R^3$, which is phenomenologically not acceptable. However, such neutrino can easily get a TeV Dirac mass with some singlet fermion from the $G_{2^31}\to G_{\rm SM}$ breaking. That is why we introduce the singlet $N_L$ in \cref{tab:fieldcontent}. If the extended-hyper-color physics generates the four-fermion operator
\begin{equation}
\mathcal{L} \supset \frac{1}{\Lambda_{N}^2}(\bar N_L \ell_{R,\alpha}^3)\left[\lambda_{N}\delta^{\alpha\beta}  (\bar \zeta^{(R)}_{R,\beta} \zeta^{(R)}_{L,1})+
\lambda^{\prime}_{N}\epsilon^{\alpha\beta} (\bar \zeta^{(R)}_{L,2} \zeta^{(R)}_{R,\beta})\right]+{\rm h.c.},\label{eq:4FLambdanu3}
\end{equation}
the chiral effective action develops
\begin{equation}
\mathcal{L} \supset \frac{c f^2\Lambda_{\rm HC}}{\Lambda_{N}^2}\left[(\bar N_L \nu_R^3) (\lambda_{N}U_{33}+ \lambda_{N}^{\prime}U^{\dagger}_{44})+
(\bar N_L \tau_R) (\lambda_{N}U_{34}-\lambda_{N}^{\prime}U^{\dagger}_{34})
\right]+{\rm h.c.}\label{eq:Nu3Mass}
\end{equation}
which gives the mass term for the neutrino $M_{N}\bar N_L\nu_R^3$, with $M_{N}= c (\lambda_{N} +\lambda^{\prime}_{N} )f\Lambda_{\rm HC}^2/g_*\Lambda_{N}^2$. This neutrino, $\nu^3_R$, will not play a role in the mass generation of the SM neutrinos.

Thus, four-fermion interactions can only generate Yukawa couplings for the third-family fermions. However, although in a suppressed way, light families have to couple to the Higgs boson too. To generate light Yukawa couplings in this TeV description, we need to go to six-fermion operators. Postponing the neutrino sector for now, the gauge symmetry allows for the following operators with two SM fermions:
\begin{align}
\mathcal{L} & \supset  \frac{1}{\tilde \Lambda_{u}^5}  \sum_{\substack{i=1,2 \\ j=1,2,3}}
(\bar q^i_{L,\alpha } u_R^j)
\sum_{\rho=1}^4
\left[\lambda^{\rho}_{u,ij} \delta^{\alpha \beta} (\bar \zeta_{L,1}^{(R)} \zeta_{R}^{\rho}) (\bar \zeta_{R}^{\rho} \zeta_{L,\beta}^{(L)}) 
+ \lambda^{\prime \rho}_{u,ij} \epsilon^{\alpha \beta} (\bar \zeta_{L,\beta}^{(L)} \zeta_{R}^{\rho}) (\bar \zeta_{R}^{\rho} \zeta_{L,2}^{(R)})\right] \nonumber\\ 
&+ 
 \frac{1}{\tilde \Lambda_{d}^5}  \sum_{\substack{i=1,2 \\ j=1,2,3}}
(\bar q^i_{L,\alpha } d_R^j)
\sum_{\rho=1}^4
\left[\lambda^{\rho}_{d,ij} \delta^{\alpha \beta} (\bar \zeta_{L,2}^{(R)} \zeta_{R}^{\rho}) (\bar \zeta_{R}^{\rho} \zeta_{L,\beta}^{(L)}) 
+ \lambda^{\prime \rho}_{d,ij} \epsilon^{\alpha \beta} (\bar \zeta_{L,\beta}^{(L)} \zeta_{R}^{\rho}) (\bar \zeta_{R}^{\rho} \zeta_{L,1}^{(R)})\right] \nonumber\\ 
&+ 
\frac{1}{\tilde \Lambda_{e}^5}  \sum_{\substack{i=1,2,3 \\ j=1,2}}
(\bar \ell^i_{L,\alpha } e_R^j)
\sum_{\rho=1}^4
\left[\lambda^{\rho}_{e,ij} \delta^{\alpha \beta} (\bar \zeta_{L,2}^{(R)} \zeta_{R}^{\rho}) (\bar \zeta_{R}^{\rho} \zeta_{L,\beta}^{(L)}) 
+ \lambda^{\prime \rho}_{e,ij} \epsilon^{\alpha \beta} (\bar \zeta_{L,\beta}^{(L)} \zeta_{R}^{\rho}) (\bar \zeta_{R}^{\rho} \zeta_{L,1}^{(R)})\right],
\label{eq:4ferLYuk}
\end{align}
where $\rho$ runs over fundamental $SU(4)$ indices, with $\rho=1,2$ corresponding to fundamental $SU(2)_L$ indices, and $\rho=3,4$ to fundamental $SU(2)_R$ indices. For instance, $\zeta^{\rho}_{L}=(\zeta^{(L)}_{L,1},\zeta^{(L)}_{L,2},\zeta^{(R)}_{L,1},\zeta^{(R)}_{L,2})$.  Due to the gauge symmetry, $\lambda^{(\prime)\rho}=(\lambda^{(\prime)L},\lambda^{(\prime)L},\lambda^{(\prime)R},\lambda^{(\prime)R})$, and the scales $\tilde\Lambda^5_{u,d,e}$ represent effective scales that can arise as product of several scales.
The chiral effective Lagrangian acquires the terms
\begin{align}
\mathcal{L} & \supset  \frac{c^{\prime} f \Lambda_{\rm HC}^5}{g_*^3\tilde \Lambda_{u}^5}  \sum_{\substack{i=1,2 \\ j=1,2,3}}
(\bar q^i_{L,\alpha } u_R^j)
\sum_{\rho=1}^4
\left[ 
\lambda^{\rho}_{u,ij} 
\delta^{\alpha \beta}
U_{\rho 3}^{\dagger} U_{\beta \rho}
+ \lambda^{\prime\rho}_{u,ij}  \epsilon^{\alpha \beta} 
U_{\rho \beta}^{\dagger} U_{4 \rho}
\right] \nonumber\\ 
&+ 
 \frac{c^{\prime}f \Lambda_{\rm HC}^5}{g_*^3\tilde \Lambda_{d}^5}  \sum_{\substack{i=1,2 \\ j=1,2,3}}
(\bar q^i_{L,\alpha } d_R^j)
\sum_{\rho=1}^4
\left[\lambda^{\rho}_{d,ij} \delta^{\alpha \beta} U^{\dagger}_{\rho 4}U_{\beta\rho} 
+ \lambda^{\prime \rho}_{d,ij} \epsilon^{\alpha \beta}
U^{\dagger}_{\rho \beta} U_{3\rho}\right] \nonumber\\ 
&+ 
\frac{c^{\prime}f \Lambda_{\rm HC}^5}{g_*^3\tilde \Lambda_{e}^5} \sum_{\substack{i=1,2,3 \\ j=1,2}}
(\bar \ell^i_{L,\alpha } e_R^j)
\sum_{\rho=1}^4
\left[\lambda^{\rho}_{e,ij} \delta^{\alpha \beta} 
U^{\dagger}_{\rho 4}U_{\beta\rho} 
+ \lambda^{\prime \rho}_{e,ij} \epsilon^{\alpha \beta}
U^{\dagger}_{\rho \beta} U_{3\rho}
\right],\label{eq:LightYuk}
\end{align}
where $c^{\prime}=O(1)$. Expanding the pNGB matrix $U$ we obtain both light-family and heavy-light mixing Yukawa couplings, that are suppressed by ${\Lambda_{\rm HC}^5}/{\tilde \Lambda_{u,d,e}^5 g_*^3} $. 
Depending on their microscopic origin, the scales $\tilde \Lambda_{u,d,e}$ can appear as a geometric average of the scales $\Lambda_{t,b,\tau}$ and higher scales, for instance: $\tilde \Lambda_{u,d,e}^5\sim \Lambda_{u,d,e} \Lambda_{t,b,\tau}^{4}$, raising the scale of the UV completion of these 6-fermion operators and protecting light-flavor observables. An example is provided in \cref{sec:EHC}.
Notice too that the Wilson coefficients $\lambda^{(\prime)\rho}$ are the product of several elementary couplings and they could partially cancel the suppression $g_*^3$ (their coupling dimension is $g^4$). In the most optimistic case, the overall suppression on light Yukawas could be $\Lambda_{\rm HC}/\Lambda_{u,d,e}$. 
We anyway emphasize that this would bring this effective description to its edge of validity, so we take this analysis as an heuristic argument to illustrate how the flavor hierarchies emerge. The precise UV description falls beyond the scope of the paper.
It is also worth mentioning that our TeV model cannot address the hierarchy between first and second families so a dynamical explanation is expected to arise at the higher scales $\Lambda_{u,d,e}$. This model can naturally appear as the IR limit of some richer UV structure with several increasing NP scales.

\subsection{Neutrino sector}
\label{sec:Neutrinos}

In \cref{tab:fieldcontent}, we have added two RH netrinos $\nu_R^{1,2}$, singlets of the full group. Their presence is natural as they could come from hypothetical UV completions of $U(1)_X$ to $SU(2)^{\prime}_R \times U(1)_{B-L}$. With these RH neutrinos, gauge symmetry allows for these six-fermion interactions
\begin{equation}
\mathcal{L} \supset 
\frac{1}{\tilde \Lambda_{\nu}^5}  \sum_{\substack{i=1,2,3 \\ j=1,2}}
(\bar \ell^i_{L,\alpha } \nu_R^j)
\sum_{\rho=1}^4
\left[\lambda^{\rho}_{\nu,ij} \delta^{\alpha \beta} (\bar \zeta_{L,1}^{(R)} \zeta_R^{\rho}) (\bar \zeta^{\rho}_R \zeta_{L,\beta}^{(L)}) 
+ \lambda^{\prime \rho}_{\nu,ij} \epsilon^{\alpha \beta} (\bar \zeta_{L,\beta}^{(L)} \zeta_R^{\rho}) (\bar \zeta^{\rho}_R \zeta_{L,2}^{(R)})\right],\label{eq:6FermNeu}
\end{equation}
giving rise to
\begin{equation}
\mathcal{L} \supset 
\frac{c^{\prime} f \Lambda_{\rm HC}^5}{g_*^3\tilde \Lambda_{\nu}^5}  \sum_{\substack{i=1,2,3 \\ j=1,2}}
(\bar \ell^i_{L,\alpha } \nu_R^j)
\sum_{\rho=1}^4
\left[\lambda^{\rho}_{\nu,ij} \delta^{\alpha \beta} 
U^{\dagger}_{\rho 3}U_{\beta\rho} 
+ \lambda^{\prime \rho}_{\nu,ij} \epsilon^{\alpha \beta}
U^{\dagger}_{\rho \beta} U_{4\rho}
\right].
\end{equation}
They generate Yukawa couplings suppressed by ${\Lambda_{\rm HC}^5}/{\tilde \Lambda_{\nu}^5 g_*^3} $ when $U$ is expanded into the pNGBs.
The smallness of the netrinos masses, $m_{\nu}\lesssim 0.1\,$eV~\cite{Planck:2018vyg}, can be achieved if $\tilde \Lambda_{\nu} \sim 10^2\,\Lambda_{\rm HC}$, in which case, neutrinos are Dirac-type. Alternatively, if $\tilde \Lambda_{\nu}\sim \tilde \Lambda_{e,u,d}\gtrsim \Lambda_{\rm HC}$ generating Yukawa couplings of the same order of magnitude than light-family Yukawas, we can implement a type-I seesaw adding large Majorana masses to $\nu_R^{1,2}$, making the active neutrinos of Majorana-type.
In any case, because the extended gauge interactions are $U(3)_{\ell}$ symmetric, the neutrino mass eigenstates are naturally misaligned to the charged lepton mass eigenstates, resulting trivially in anarchic mixing angles of the PMNS matrix.

This version of the model has two massive and one massless active neutrinos, which is compatible with current observations. However, it is always possible to add a third mass eigenstate by including another singlet $\nu_R$.\footnote{We could also add a small Majorana mass to $N_L$, implementing an inverse seesaw mechanism with $N_L$, $\nu_R^3$ and $\nu^3_L$. However, due to $SU(2)_R^{\ell_3}$, the Yukawa coupling between $\nu_L^3$ and $\nu_R^3$ in~\cref{eq:YukawasH3} is aligned to the $\tau$, and it would introduce some alignment in the PMNS matrix. Still, given the misalignment introduced by $\nu_R^{1,2}$, it would be possible to reproduce the SM PMNS if the contribution from this inverse seesaw mechanism is smaller. Anyway, we choose not to implement this to avoid any alignment. Notice that in the alternative model with a gauged $U(1)_{R2}$ instead of the full $SU(2)_{R2}$, implementing this inverse seesaw does not introduce any alignment in the PMNS matrix.\label{foot:U1R0}}

\subsection{Scalar potential}
\label{sec:Scalar}

A fundamental aspect of every composite Higgs model is to ensure the Higgs boson develops the SM potential. 
Every breaking of the global symmetry $SU(4)_1\times SU(4)_2$ gives a contribution to the pNGB potential. 
Like the Yukawa couplings, these contributions can be estimated using again the transformation properties of the spurions or breaking coefficients to build $SU(4)_1\times SU(4)_2$ invariant terms, taking into account possible loop factors, and inserting the appropriate powers of $\Lambda_{\rm HC}$ and $g_*$ to adjust the mass and coupling dimensions.
These breakings include the gauging and the extended sector that generates the Yukawa couplings, mainly to the third generation (the others are suppressed). 
Other breakings usually considered are hyper-quark masses~\cite{Ma:2015gra}, in the same way quark masses contribute to the pion masses. However in our model, the gauge symmetry forbids any mass term between the hyper-quarks.

\subsubsection{pNGB masses}

We start discussing the different contributions to the pNGB masses
\begin{equation}
V_{\rm scalar}\supset \frac{1}{2} m_S^2 S^2+m_{H_1}^2|H_1|^2+m_{H_2}^2|H_2|^2+
(m_{12}^2\, H^{\dagger}_1 H_2+{\rm h.c.}).\label{eq:pNGBMasses}
\end{equation}
Loops of non-hyper-colored particles contribute to the potential through their symmetry-breaking couplings to the hyper-sector. 
The one-loop contributions of the deconstructed EW gauge bosons vanish at order $O(g^2)$ in the gauge couplings:
each gauge coupling only breaks one of the $SU(4)$ factors of the global group $SU(4)_1\times SU(4)_2$. However, to generate a potential we need to break both $SU(4)_1$ and $SU(4)_2$, so the leading non-vanishing contributions are $O(g_1^2g_2^2)$, where  $g_{1,2}$ are the gauge couplings on the sites $1$ and $2$: $g_1=g_{L1},g_{X}$ and $g_2=g_{L2},g_{R2}$. Schematically,
\begin{align}
V_{\rm scalar} \supsetsim   -\frac{f^4}{16\pi^2}g_{1}^2g_{2}^2\, {\rm Tr} (U^{\dagger}T^a_1 UT^b_2){\rm Tr} (U^{\dagger}T^a_1 UT^b_2),
\end{align}
with $T^a_{1,2}$ corresponding to the generators associated to the gauge couplings $g_{1,2}$, and $a,b$ the adjoint gauge boson indices. The pNGB mass contribution is then $O(g_1 g_2 f/4\pi)$.
This is essentially the idea behind the concept of collective breaking exploited in little Higgs models to suppress leading contributions to the Higgs potential~\cite{Arkani-Hamed:2001nha,Arkani-Hamed:2002iiv,Arkani-Hamed:2002ikv,Schmaltz:2005ky,Perelstein:2005ka,Schmaltz:2010ac,Chung:2023gcm}
and it is a consequence of the Higgs boson emerging from the breaking of the deconstructed gauge symmetry.

Four-fermion operators of~\cref{eq:ModelI4ferYuk} that generate the Yukawa couplings of the third-family SM fields contribute at one loop with leading terms
\begin{align}
V_{\rm scalar}\supset&-\frac{a_f f^4}{16\pi^2} \bigg[
N_c\frac{\Lambda_{\rm HC}^4}{\Lambda_t^4} \left|\lambda_t U^{\dagger}_{\alpha 3}+\lambda_t^{\prime}
\epsilon_{\alpha \alpha^{\prime}} U_{4 \alpha^{\prime}}\right|^2
+
N_c\frac{\Lambda_{\rm HC}^4}{\Lambda_b^4} \left|\lambda_b U^{\dagger}_{\alpha 4}+\lambda_b^{\prime}
\epsilon_{\alpha \alpha^{\prime}} U_{3 \alpha^{\prime}}\right|^2
\nonumber\\
&+
\frac{\Lambda_{\rm HC}^4}{\Lambda_{\tau}^4} \left|\lambda_{\tau}U^{\dagger}_{\alpha (\beta+2)}+\lambda_{\tau}^{\prime}
\epsilon_{\beta \beta^{\prime}}\epsilon_{\alpha\alpha^{\prime}} U_{(\beta^{\prime}+2) \alpha^{\prime}}\right|^2
+
\frac{\Lambda_{\rm HC}^4}{\Lambda_{N}^4} \left|\lambda_{N}U_{3(\alpha+2)}+\lambda_{N}^{\prime}
\epsilon_{\alpha \alpha^{\prime}} U^{\dagger}_{(\alpha^{\prime}+2)4} \right|^2
\bigg],\label{eq:V1}
\end{align}
where $a_f=O(1)$, $\alpha^{(\prime)},\beta^{(\prime)}=1,2$, and their explicit contribution to the pNGB masses squared are given in \cref{eq:mSL,eq:mH1L,eq:mH2L,eq:m12L} of \cref{sec:AppPot}.
We can see that the top Yukawa contribution to the pNGB masses squared is negative and $O(f^2y_t^2 N_{c}/N_{\rm HC})$, too large given the phenomenological constraints on $f$ we discuss later, $f>(2-3)\,$TeV.\footnote{In little Higgs models, vector-like fermions are also added to eliminate the leading contribution of~\cref{eq:V1}. Here we explore the minimal model with the minimal fermion content.} 
However, the contribution from $\lambda_N^{(\prime)}$, necessary to give a TeV mass $M_N$ to the neutrino $\nu_R^3$ in \cref{eq:4FLambdanu3}, is positive and can partially cancel or overcome the top-quark one if $\lambda_{N}^{(\prime)}/\Lambda_N^2\sim y_t/f\Lambda_t$. According to \cref{eq:Nu3Mass}, this implies $M_N\sim y_t f$.

Apart from the masses, the coefficients $\lambda_N^{(\prime)}$ also generate a tadpole for the field~$S$,
\begin{equation}
V_{\rm scalar}\supset \frac{\sqrt{2}}{32\pi^2 }\frac{\Lambda_{\rm HC}^4}{\Lambda_{N}^4} a_f f^3\,  {\rm Im}(\lambda^{*}_N \lambda^{\prime}_N) S.\label{eq:STad}
\end{equation}
We will however assume that the extended sector is close to be CP conserving, so the Wilson coefficients $\lambda^{(\prime)}$ can be chosen approximately real and this tadpole can be neglected.\footnote{Imaginary parts of $\lambda$ similar to the real ones may generate CP violating operators in the SMEFT at dimension 6,
\begin{equation}
\mathcal{L}\supsetsim   \frac{y_t^2}{16\pi^2 \Lambda_t^2} (g_Y^2|H|^2 B^{\mu\nu} \tilde B_{\mu\nu}
+g_L^2|H|^2 W_a^{\mu\nu} \tilde W^a_{\mu\nu}
+g_L g_YH^{\dagger} \sigma^a H  W_a^{\mu\nu} \tilde B_{\mu\nu}
).
\end{equation}
They would contribute to the electric dipole moment of the electron at one loop, imposing $4\pi \Lambda_{t}/y_t$ larger than several hundreds of TeV~\cite{Kley:2021yhn}.}

Depending on the extended hyper-color sector (see \cref{sec:EHC}), on top of the effective operators already discussed carrying two SM fermions and two hyper-quarks, this sector might also generate four-hyper-quark operators:
\begin{equation}
\mathcal{L}\supset \frac{1}{\Lambda_{\zeta}^2} \sum_{\rho_i=1}^4\left[ \lambda_{\rho_1\rho_2\rho_3\rho_4}
(\bar \zeta_L^{\rho_1}\zeta_R^{\rho_2})(\bar \zeta_R^{\rho_3}\zeta_L^{\rho_4})+ \left(
\lambda^{\prime}_{\rho_1\rho_2\rho_3\rho_4}
(\bar \zeta_L^{\rho_1}\zeta_R^{\rho_2})(\bar \zeta_L^{\rho_3}\zeta_R^{\rho_4})+{\rm h.c.}\right)
\right],\label{eq:EHCt}
\end{equation}
where the gauge symmetry restricts these Wilson coefficients as specified in \cref{sec:AppPot} to 12 independent ones. They are spurions of the global symmetry $SU(4)_1\times SU(4)_2$ transforming as $\lambda\sim (\bar {\bf 4}\times {\bf 4},\bar {\bf 4}\times {\bf 4})$ and  $\lambda^{\prime}\sim ( {\bf 4}\times {\bf 4},\bar {\bf 4}\times\bar {\bf 4})$.
This is an explicit breaking of the symmetry that also contributes to the potential like:
\begin{align}
V_{\rm scalar} \supset -\frac{f^4\Lambda_{\rm HC}^2}{\Lambda_{\zeta}^2} \left[a\left( \lambda_{\rho_1 \rho_2 \rho_3 \rho_4} U^{\dagger}_{\rho_2 \rho_1}  U_{\rho_4 \rho_3} \right) + 
a^{\prime}\left( \lambda^{\prime}_{\rho_1 \rho_2 \rho_3 \rho_4} U^{*}_{\rho_1 \rho_2}  U^{*}_{\rho_3 \rho_4}+{\rm h.c.}\right)
\right],
\label{eq:V0}
\end{align}
with $a^{(\prime)}=O(1)$.
The explicit contributions to the pNGB masses are given in \cref{eq:mST,eq:mH1T,eq:mH2T,eq:m12T} of \cref{sec:AppPot}. If
\begin{equation}
\frac{\lambda_{4\zeta}}{\Lambda_{\zeta}^2} \sim \frac{N_{c}}{N_{\rm HC}} \frac{y_t^2}{\Lambda_{\rm HC}^2},
\end{equation}
with $\lambda_{4\zeta}$ generically representing the Wilson coefficients of \cref{eq:EHCt},
we obtain same-order contributions than from the top Yukawa one in~\cref{eq:V1}. These four-hyper-quark operators are not necessary to reproduce a viable SM Higgs potential, but we consider them because they appear naturally in some of the extended hyper-color sector possibilities.
The Wilson coefficients $\lambda^{\prime}$ in \cref{eq:V0} can also generate a tadpole for $S$ (given in \cref{eq:STadp}) that, like in \cref{eq:STad}, will vanish if the extended sector is CP conserving.

Before discussing the necessary little tuning of the model, we comment on the mass mixing terms  between the Higgses of~\cref{eq:pNGBMasses}. The real part can be removed by rephasing the hyper-quark fields with the appropriate phase, $\zeta^{(L)}\to e^{-i\alpha}\zeta^{(L)}$, and $\zeta^{(R)}\to e^{i\alpha}\zeta^{(R)}$. This  is the transformation $g={\rm diag}(e^{-i\alpha},e^{-i\alpha},e^{i\alpha},e^{i\alpha})$ in $SU(4)_V$, and acts over the pNGB as a real rotation of angle $2\alpha$ between $H_1$ and $H_2$~\cite{Ma:2015gra}.
After this redefinition, we can get ${\rm Re}\,m_{12}^2=0$.
However, if ${\rm Im} \,m_{12}^2 \neq 0$, once one of the Higgses, say $H_1$, develops a VEV that we can choose real, 
$\langle H_1\rangle= (0,v_1)^t$, $H_2$ will also develop another one $\langle H_2\rangle= (0,i v_2)^t$ which will break custodial symmetry as discussed below~\cref{eq:H1DH2}.
Indeed, if we want to integrate out the heaviest Higgs, we first need to rotate to the basis $(H_1^{\prime},H_2^{\prime})$ that diagonalizes the mass matrix in the EW symmetric phase:
\begin{equation}
\begin{pmatrix}
H_1 \\
H_2
\end{pmatrix} =
\begin{pmatrix}
c_H && i s_H \\
i s_H && c_H 
\end{pmatrix}
\begin{pmatrix}
H^{\prime}_1 \\
H^{\prime}_2
\end{pmatrix},\label{eq:ImRot}
\end{equation}
where $s_H=\sin (\theta_H)$, $c_H=\cos (\theta_H)$,  $H_1^{\prime}$ is the negative mass squared eigenstate and $H_2^{\prime}$ is the heaviest Higgs.
Substituting in the last line of~\cref{eq:H1DH2}, it gives a contribution to the custodial-breaking operator $\mathcal{L} \supset C_{HD}|H^{\dagger}DH|^2$,
\begin{equation}
\Delta C_{HD}=-\frac{c_H^2 s_H^2}{2f^2},
\end{equation}
which for an $O(1)$ mixing angle, gives the experimental constraint $f\gtrsim 3\,$TeV at $95\%\,$C.L. in absence of any other effect.
We will see in \cref{sec:Pheno} that there are other effects to take into account, and an $O(1)$ mixing angle between the two Higgses actually helps phenomenologically because it partially cancels other contributions.
This imaginary mass mixing depends on the physics realizing the four-fermion operators as we can see in \cref{sec:AppPot,sec:EHC}. For instance, it can vanish if the four-fermion operators are generated by scalar fields, bi-doublets of the $SU(2)_{Xi}$ gauge groups.

Regarding the pNGB mass eigenvalues, to reproduce correctly the Higgs mass, we will assume that the different contributions to one of Higgs eigen-masses almost cancels out, leaving a final mass $m_{H_1^{\prime}}^2\sim -\lambda_H v^2/2$ while the other Higgs and $S$ get a positive mass squared $O(f^2)$. Assuming $f\sim (2-3)\,$TeV (see \cref{sec:Pheno}), the tuning necessary to achieve this is at the $10^{-2}-10^{-3}$ level. 
For instance, assuming a benchmark such that the only Wilson coefficients relevant for the pNGB potential are $\lambda_t$ and $\lambda_N=\lambda_N^{\prime}$, we obtain
\begin{equation}
\theta_H=\frac{\pi}{4},~m^2_{H^{\prime}_1}\sim \frac{M_N^2-12f^2 y_t^2}{4N_{\rm HC}},~m^2_{H^{\prime}_2}=2m_S^2\sim\frac{M_N^2}{4N_{\rm HC}},\label{eq:benchmark}
\end{equation}
where the tuning requires $M_N^2 \lessapprox 12 f^2 y_t^2  $.

%%%%

\subsubsection{Higgs quartic}

Loops of non-hyper-colored fermions in \cref{eq:V1} and four-hyper-quark operators in \cref{eq:V0} can also generate a viable Higgs quartic only for $O(1)$ imaginary mixing angles in~\cref{eq:ImRot}. 
In the decoupling limit $m_{H_2^{\prime}}\gg v$ we have $v_1 =c_H v +O(v^2/m_{H_2^{\prime}}^2)$ and $v_2 =s_H v+O(v^2/m_{H_2^{\prime}}^2)$. Then, in \cref{eq:H2x2,eq:UMatrix},
\begin{equation}
\langle H_{2\times 2} \rangle =
\frac{v}{\sqrt{2}}
\begin{pmatrix}
c_H+s_H && 0 \\
0 && c_H-s_H 
\end{pmatrix},
\end{equation}
and the matrix $U$ in the vacuum takes the form
 \begin{equation}
U=
\left(
\begin{array}{cc:cc}
\cos[(c_H+ s_H)\xi] & 0 & \sin[(c_H+ s_H)\xi]  & 0 \\
0 & \cos[(c_H- s_H)\xi] & 0 & \sin[(c_H- s_H)\xi]  \\
\hdashline
- \sin[(c_H+ s_H)\xi] &0 & \cos[(c_H+ s_H)\xi] & 0   \\
0 & - \sin[(c_H- s_H)\xi] & 0 & \cos[(c_H- s_H)\xi]    \\
\end{array}
\right),
\end{equation}
with $\xi = v/(2\sqrt{2} f)$. Inserting $U$ into \cref{eq:V1,eq:V0}, several independent trigonometric functions appear,
\begin{align}
 V&_{\rm scalar}/f^4 \supset \alpha_+ \sin^2[(c_H+ s_H)\xi]+\alpha_-\sin^2[(c_H- s_H)\xi]\nonumber\\
&+\alpha_{\rm sin}\sin[(c_H+ s_H)\xi] \sin[(c_H- s_H)\xi]+\alpha_{\rm cos}\cos[(c_H+s_H)\xi] \cos[(c_H- s_H)\xi].\label{eq:VwithsH}
\end{align}
These parameters $\alpha$ are given in \cref{eq:alphap1,eq:alpham1,eq:alphas1,eq:alphac1,eq:alphap0,eq:alpham0,eq:alphas0,eq:alphac0} of \cref{sec:AppPot}.
The potential of \cref{eq:VwithsH} can reproduce the SM Higgs potential with a suppressed mass compared to $f$ and the correct quartic if $s_H=O(1)$. 
For instance, the benchmark of \cref{eq:benchmark} would give 
\begin{equation}
\lambda_H\sim  \frac{3  y_t^2 }{8N_{\rm HC}},
\end{equation}
with $\mathcal{L}\supset \lambda_H |H|^4/2$, which gives a very acceptable quartic for not too large $N_{\rm HC}$.
However, for vanishing or suppressed mixing angle, $s_H\approx 0$, all the terms of \cref{eq:VwithsH} only contribute to $V_{\rm scalar} = \alpha f^4  \sin^2 \xi$. Once we tune $\alpha \sim -0.04 \, {\rm TeV}^2/f^2$ to reproduce SM Higgs mass, the quartic becomes very suppressed and the Higgs develops a VEV at the scale $f$.
To bring down the Higgs VEV to the EW scale in this case, we need an independent trigonometric function, $\Delta V_{\rm scalar}  =\beta f^4 \sin^4 \xi$, providing independent contributions to the Higgs quartic (to reproduce the SM, $\beta \sim 1$). Higher-order terms from the 
four-fermion operators, schematically
\begin{equation}
V_{\rm scalar} \supsetsim
 \frac{N_c f^4}{16 \pi^2g_*^4}\sum_{f=t,b,\tau,N} \frac{\Lambda_{\rm HC}^8}{\Lambda_{f}^8}|\lambda_f  U+\lambda_f^{\prime}  U|^4
 + \frac{f^4\Lambda_{\rm HC}^4}{\Lambda_{\zeta}^4} 
 \frac{1}{g_*^2}[\lambda_{4\zeta} \lambda_{4\zeta} U U U U],\label{eq:VHigOr}
\end{equation}
can give contributions to $\beta$ order $ \sim \,N_c \,y_t^4/16\pi^2$.
Another interesting class of four-hyper-quark operators are
\begin{equation}
\mathcal{L}\supset \frac{1}{\Lambda_{\zeta}^2} \sum_{\rho_i=1}^4\left[ \lambda^{(1)}_{\rho_1\rho_2\rho_3\rho_4}
(\bar \zeta_L^{\rho_1}\gamma_{\mu}\zeta_L^{\rho_2})(\bar \zeta_L^{\rho_3}\gamma^{\mu}\zeta_L^{\rho_4})
+\lambda^{(2)}_{\rho_1\rho_2\rho_3\rho_4}
(\bar \zeta_R^{\rho_1}\gamma_{\mu}\zeta_R^{\rho_2})(\bar \zeta_R^{\rho_3}\gamma^{\mu}\zeta_R^{\rho_4})
\right],\label{eq:EHCt2}
\end{equation}
where gauge symmetry again restricts the possible Wilson coefficients $\lambda^{(1,2)}_{\rho_1\rho_2\rho_3\rho_4}$. They implement a collective breaking similar to the gauge breaking: they individually break only one of the $SU(4)$ factors of the global group, so the potential only gets contributions at order $O(\lambda^{(1)}\lambda^{(2)})$,
\begin{align}
V_{\rm scalar} \supsetsim  \frac{f^4\Lambda_{\rm HC}^4}{g_*^2\Lambda_{\zeta}^4}[\lambda^{(1)} \lambda^{(2)} U^{\dagger} U U^{\dagger} U] .\label{eq:V0B}
\end{align}
The point of these operators is that they contribute to the Higgs quartic like
\begin{equation}
\Delta \beta \sim \frac{\lambda^{(1)} \lambda^{(2)}}{g_*^2}\frac{ \Lambda_{\rm HC}^4}{\Lambda_{\zeta}^4},
\end{equation}
without giving large corrections to the masses, so they could be necessary in case the higher-orders of~\cref{eq:VHigOr} are not enough to achieve the size of the SM Higgs quartic in the case $s_H\approx 0$.

%%%%

\subsection{Down-type Yukawa couplings}
\label{sec:Down}

Concerning the $b$ and $\tau$ Yukawa couplings, several scenarios could address their hierarchically suppressed values:
\begin{enumerate}
\item[(1)] The Yukawas $y_{b,\tau}$ are originated at higher scales $\Lambda_{b,\tau}>\Lambda_t$, which explains their suppression: $y_{b,\tau}/y_t \sim \Lambda_t^2/\Lambda_{b,\tau}^2$. 
\item[(2)] The scales responsible of the down-type Yukawa couplings are similar to the top-quark one, $\Lambda_{b,\tau}\sim \Lambda_t$, but the Higgs VEVs are hierarchical: let $H_b$ be the only linear combination of the Higgses that couple to down-type fields, and $H_t$ the perpendicular one, with an $O(1)$ coupling to the top quark. The top-bottom hierarchy then appears because $v_t\gg v_b$.  This is similar to the mechanism suggested in~\cite{Rosenlyst:2020znn}.
We can implement this idea in different ways:
\begin{enumerate}
\item[(a)] The four-fermion operators are such that $\Delta m_{12}^2=0$, $\lambda_b^{\prime}=\lambda_b^{\prime}$ and $\lambda_{\tau}=-\lambda_{\tau}^{\prime}$. This can be realized if they are generated by scalar fields, bi-doublets of the $SU(2)_{Xi}$ gauge groups (see \cref{sec:AppPot,sec:EHC}).
Then, the Yukawa couplings of $b$ and $\tau$ to $H_1$ vanishes, but not to $H_2$, so $H_t=H_1$ and $H_b=H_2$. At leading order, $H_1$ gets an EW VEV but $H_2$ has a positive mass order $f$. At a higher scale $\Lambda^{\prime}$, with $y_{b,\tau}/y_t \sim \Lambda_t^2/\Lambda^{\prime\,2}$, an extended sector generates an imaginary and suppressed mass mixing $\sim y_{b,\tau}/y_t f^2$. This mass mixing will induce a $y_{b,\tau}/y_t$-suppressed VEV to $H_2$, generating the appropriate mass for the $b$ and $\tau$.
\item[(b)] The NP implementing the four-fermion operators only generates the Wilson coefficients $\lambda_{t,b,\tau}$ and $\lambda^{\prime}_{t,b,\tau}=0$ (see \cref{sec:EHC} for examples of this). Then, $H_t=(H_1-iH_2)/\sqrt{2}$ and $H_b=(H_1+iH_2)/\sqrt{2}$. If $\theta_H \approx \pi/4$ in~\cref{eq:ImRot}, implying $\langle H_1 \rangle \approx (0,v/\sqrt{2})^t $ and $\langle H_2 \rangle \approx (0,i v/\sqrt{2})^t $, the final $b$ and $\tau$ Yukawa couplings in \cref{eq:YukawasH3} are suppressed with respect to the top-quark one.
\end{enumerate}
\end{enumerate}

\section{Phenomenology}
\label{sec:Pheno}

The phenomenology of our model is mostly dictated by the new states appearing at the scale $f$: the new massive vector bosons and the second Higgs. The extra scalar singlet $S$ decays into EW gauge bosons via Wess-Zumino-Witten (WZW) terms~\cite{Wess:1971yu,Witten:1983tw}, in a similar way than the neutral pion decays into photons, $\pi^0\to \gamma \gamma$~\cite{Adler:1969gk,Bell:1969ts}, as it is shown in~\cite{Ma:2015gra}, and does not play a dominant role in the phenomenological analysis.
Also, the heavy neutral lepton (HNL) formed by the pair $\nu_R^3$ and $N_L$ develops a tiny mixing angle with the $\tau$ neutrino, $\theta_{\tau}=m^2_{\tau}/M_N^2$, due to the Yukawa coupling between $\nu^3_R$ and $\ell_L^3$ of \cref{eq:ModelI4ferYuk}, which is the same than the $\tau$-lepton one because of $SU(2)_{R}^{\ell_3}$. HNLs mixing with the $\tau$ neutrino above the EW scale are mainly constrained by EW physics and lepton flavor universality (LFU) tests in $\tau$ decays so $\theta_{\tau}\leq 0.05$ at $95\%$~C.L. \cite{Lizana:2023kei}. We see that such a bound is irrelevant for us due to the small Yukawa coupling.
Thus, LHC searches of the new vector bosons, EW precision observables (EWPO) affected by the new vector bosons and the second Higgs, and flavor observables sensitive to the breaking of universality, are the main observables to study.
In \cref{sec:ApSMEFT} we show the dimension 6 Wilson coefficients of the SM effective field theory (SMEFT) generated after integrating out the new vector bosons and from the expansion of the leading term of the CCWZ action in~\cref{eq:H1DH2}, after the rotation of \cref{eq:ImRot} and reducing to the Warsaw basis~\cite{Grzadkowski:2010es,Gherardi:2020det}.

Other effects typically studied in composite Higgs models come from linear mixings between the EW gauge bosons and heavy composite resonances, generating dimension 6 operators such as~\cite{Giudice:2007fh},
\begin{equation}
\mathcal{L} \supset \frac{{\cal C}_W g_L}{2g_*^2 f^2} 
(H^{\dagger} \sigma^a D_{\nu} H) \nabla_{\mu}
W_a^{\mu\nu} + 
 \frac{{\cal C}_B g_Y}{2g_*^2 f^2} 
(H^{\dagger} D_{\nu} H) \partial_{\mu} B^{\mu\nu}.
\end{equation}
Depending on the specific realization of the extended hyper-color sector (see \cref{sec:EHC}), operators of the form
\begin{equation}
\mathcal{L}\supset
\frac{{\cal C}_{Hq}^{(3)}}{g_* f^2} (H^\dagger  \sigma_a  i\overleftrightarrow{D}_{\mu} H)\,  \bar q_L^3\sigma^a \gamma^{\mu} q^3_L + \frac{{\cal C}_{Ht}}{g_* f^2} (H^\dagger i \overleftrightarrow{D}_{\mu} H)\, \bar t_R\gamma^{\mu} t_R\label{eq:HqOp}
\end{equation}
could also appear, affecting the $Z$ coupling to $b$-quarks, and running into universal effects in the EWPO.
The extended sector can also generate third-generation four-quark operators,
\begin{align}
\mathcal{L}\supset &
\,\,\frac{{\cal C}_{qq}^{(1)}}{g_*^2 f^2}  (  \bar q_L^3 \gamma_{\mu} q^3_L)  (  \bar q_L^3 \gamma^{\mu} q^3_L)+
\frac{{\cal C}_{qq}^{(3)}}{g_*^2 f^2}  (  \bar q_L^3\sigma^a \gamma_{\mu} q^3_L)  (  \bar q_L^3\sigma_a \gamma^{\mu} q^3_L)
\nonumber\\
& + \frac{{\cal C}_{tt}}{g_*^2 f^2}  (\bar t_R^3\gamma_{\mu} t_R) (\bar t_R^3\gamma^{\mu} t_R) 
+\frac{{\cal C}^{(1)}_{qt}}{g_*^2 f^2}  (  \bar q_L^3 \gamma_{\mu} q^3_L)   (\bar t_R\gamma^{\mu} t_R),\label{eq:qqOp}
\end{align}
running into $Z\to bb$, and, at order $({\rm log})^2$, into universal effects in the EWPO~\cite{Allwicher:2023aql,Stefanek:2024kds}.
The $O(1)$ coefficients controlling these effects,  ${\cal C}$, depend on the strong dynamics and the specific extended hyper-sector, 
so they cannot be fixed by symmetry arguments and are unknown at this level. Anyway, their effects are suppressed by $g_*=4\pi/\sqrt{N_{\rm HC}}$. For not too large values of $N_{\rm HC}$, we expect them to only give a subleading correction to the limits we will find from integrating out the massive vector bosons and we neglect here their effects.

\subsection{LHC searches}

One of the main phenomenological consequences of performing a TeV deconstruction of the SM gauge group is the appearance of TeV vector bosons that could be observed at LHC. The search of vector bosons~\cite{deBlas:2012qp,Pappadopulo:2014qza} will set one of the main limits of the scale $f$ of our model.
For $O(1)$ mixing angles $\theta_L$ and $\theta_X$ of \cref{eq:thetaLX}, our model will be strongly constrained by $pp\to {\cal B}_0, {\cal W}^0\to \ell^+ \ell^-$ and $pp\to {\cal W}^{\pm}\to \ell \nu$ searches, where here $\ell=e,\mu$. For vector bosons with couplings to light fermions of the order of the EW gauge couplings or larger, these limits are dominated by non-resonant searches.\footnote{As an illustration, an $SU(2)_L$ triplet with a universal coupling to all SM LH fermions with value $0.54$ (model A of~\cite{ATLAS:2024qvg}) has a limit in resonant searches of $5.8$\,TeV at $95\,\%$ C.L., but of $7.4$\,TeV in non-resonant searches (obtained with \texttt{HighPT}).} Therefore, we construct a $ \chi^2_{pp\to\ell \ell, \ell \nu}$ from $pp\to \ell^+ \ell^-$ and $pp\to \ell \nu$ precesses using \texttt{HighPT}~\cite{Allwicher:2022mcg,Allwicher:2022gkm} and the semileptonic operators resulting from integrating out $\cal W$ and ${\cal B}_0$ given in \cref{sec:ApSMEFT}. For small mixing angles $\theta_L$ and $\theta_X$, ${\cal W}$ and ${\cal B}_0$ decouple from light quarks and light leptons, making these searches irrelevant.
Regarding the triplet $\cal W$, for small $\theta_L$ it will couple mostly to $q_3$. Combined searches of the vector triplet in all the channels~\cite{ATLAS:2024qvg} show that when the coupling to $q_3$ is large and there are no couplings to leptons, the limits are strongly dominated by $pp\to {\cal W}^{\pm} \to t b$. We then extract from the ATLAS search with $139\,$fb$^{-1}$ in Ref.~\cite{ATLAS:2023ibb} the limits of a universal ${\cal W}^{\pm}$ on $\sigma(pp\to {\cal W}^{\pm}_{\rm Univ.}\to bt)$ as function of the triplet mass $M_{\cal W}$ and construct our $\chi^2$,
\begin{equation}
\chi^2_{pp\to bt} = 4 \left(\frac{\sigma(pp\to {\cal W}^{\pm}\to bt;M_{\cal W}) }{\sigma_{95\%}(pp\to {\cal W}^{\pm}_{\rm Univ.}\to bt;M_{\cal W})}\right)^2,
\end{equation}
where $\sigma_{95\%}$ is the extracted limit at $95\%$ C.L.
Thus, it depends quadratically in the cross section, i.e. the observable, and it reproduces the correct $95\,\%$ C.L. limit.
Similarly, for small $\theta_X$, ${\cal B}_0$ is mostly coupled to $\tau_R$ and searches of $pp\to Z^{\prime}\to \tau^+\tau^-$ become relevant. We extract the limits of a sequential $Z^{\prime}_{\rm SSM}$ in this channel from the CMS search with $138\,$fb$^{-1}$ in Ref.~\cite{CMS:2024pjt}. We then construct a $\Delta \chi^2_{pp\to \tau\tau}$ as before:
\begin{equation}
\chi^2_{pp\to \tau\tau} = 4 \left(\frac{\sigma(pp\to {\cal B}_0\to \tau^+\tau^-;M_{{\cal B}_0}) }{\sigma_{95\%}(pp\to {Z^{\prime}_{\rm SSM}}\to \tau^+\tau^-;M_{{\cal B}_0})}\right)^2.
\end{equation}
The parton luminosity functions to calculate our cross sections are taken from \texttt{ManeParse}~\cite{Clark:2016jgm}.
The vector ${\cal B}_1$ can only be produced via vector-boson fusion, decreasing significantly its production cross section. Furthermore, if $\nu_3$ is heavier than ${\cal B}_1$, something that naturally can happen, this vector will only decay to SM bosons, ${\cal B}^{\pm}_1\to W^{\pm}Z$ and ${\cal B}^{\pm}_1\to W^{\pm}h$~\cite{Grojean:2011vu}. Its search therefore provides limits significantly weaker that we can neglect here.
We also expect anomalous vertices between the new and the EW vector bosons from loops of hyper-quarks due to WZW terms (explicitly built for this type of models in Ref.~\cite{Fuentes-Martin:2024fpx}). However, we also expect these effects to be phenomenologically subdominant, with the couplings to fermions that we study here playing the leading role.

LHC searches are then combined into a single $ \chi^2_{\rm LHC}$. Since every $\chi^2$-function corresponds to different channels, we can sum them:
\begin{equation}
\chi^2_{\rm LHC}=\chi^2_{pp\to \ell\ell,\ell\nu}+ \chi^2_{pp\to bt}+ \chi^2_{pp\to \tau\tau}.
\end{equation}

We will also show the projected limits assuming no BSM signal is found for the high-luminosity phase of LHC (HL-LHC). We use \texttt{HighPT} to extract the projection of $\Delta\chi^2_{pp\to \ell \ell,\ell\nu}$ for an integrated luminosity of $3\,$ab$^{-1}$.
For $\chi^2_{pp\to tb}$ and $\chi^2_{pp\to \tau \tau}$, we rescale the extracted $95\,\%\,$C.L. limits on the cross sections. Because both signal and background scale linearly in the luminosity $L$, and statistical relative fluctuations decrease like $1/\sqrt{L}$, the limits on the NP signal cross section typically scale like $1/\sqrt{L}$ when there is background and like $1/L$ when there is not. Based on the backgrounds provided in~\cite{ATLAS:2023ibb,ATLAS:2020zms}, we estimate the rescaling factor $\sqrt{139/3000}$ for low masses, $M_{\cal W}<5.5\,$TeV and $M_{{\cal B}_0}<2\,$TeV, and $139/3000$ for high masses, $M_{\cal W}>6.5\,$TeV and $M_{{\cal B}_0}>3\,$TeV, interpolating between both extremes in both searches.

Searches of a second Higgs could also play a role~\cite{Branco:2011iw,ParticleDataGroup:2024cfk,ATLAS:2024itc}, specially if the top-bottom hierarchy is realized via hierarchical VEVs $v_t\gg v_b$, because then the second Higgs couples with $O(1)$ Yukawa couplings to $b$ and $\tau$. The mass exclusion limit in those cases is around $(1-2)\,$TeV~\cite{ATLAS:2020zms}. If $b$ and $\tau$ couplings to the second Higgs are suppressed, limits can go down below the TeV. The second Higgs mass is expected to be at the scale $f$, but it is essentially an independent parameter since it also depends on the Wilson coefficients of the multiple four-fermion operators introduced. We focus our LHC analysis on the searches to new massive vectors, that we expect to constrain the scale~$f$ more significantly.

\subsection{EW precision data}

We perform an EW fit employing the \{$\alpha_{EM} , m_Z , G_F $\} input scheme and the observables specified in Table 1 and 2 of Ref.~\cite{Breso-Pla:2021qoe}. We update the $W$ boson mass value: its SM prediction is taken to be $m_W^{\rm SM}=(80361\pm 6)\,$MeV~\cite{ParticleDataGroup:2016lqr,Bagnaschi:2022whn} and we take the updated experimental value used in~\cite{Erdelyi:2024sls}, which combines the results from LEP2, Tevatron (assuming a pre-2022 CDF measurement), LHCb~\cite{ParticleDataGroup:2022pth}, ATLAS~\cite{ATLAS:2023fsi} and CMS~\cite{CMS:2024lrd},
$
m_W^{\rm Exp}= ( 80367\pm 7)\,
$MeV.
The contributions of the NP Wilson coefficients to the EW observables, in the narrow width approximation, can be parametrized by the phenomenological Lagrangian,
\begin{align}
\mathcal{L} \supset &- \frac{g_L}{\sqrt{2}}W^{+\mu}\bigg[\bar u_L^i \gamma_{\mu} \left(V_{ij}
+\delta g_{ij}^{Wq}\right)d_L^j 
+ \bar \nu_L^i \gamma_{\mu} \left(\delta_{ij}+\delta g_{ij}^{W \ell }\right)e_L^j\bigg]+{\rm h.c.}\nonumber \\
&-\frac{g_L}{c_W}\, Z^{\mu} \bigg[ \bar \psi^i_L \gamma_{\mu}
\left( g^{Z}_{\psi_L}\delta_{ij}+ \delta g_{ij}^{Z\psi_L}\right) \psi_L^j
+\bar \psi^i_R \gamma_{\mu}
\left( g^{Z}_{\psi_R} \delta_{ij}+ \delta g_{ij}^{Z\psi_R}\right) \psi_R^j
\bigg] \nonumber \\
&+\frac{g_L^2 v^2}{4}(1+\delta m_W)^2 W^{+\mu} W^-_{\mu}+\frac{g_L^2 v^2}{8 c_W^2}Z^{\mu} Z_{\mu},
\label{EWEffLag}
\end{align}
where $\delta x$'s encode the NP contributions,
$
g^{Z}_{\psi_L}=\,T_{\psi}^3-s_W^2 Q_{\psi}$, $g^{Z}_{\psi_R}= -s_W^2 Q_{\psi}
$, and $s_W$ is the sine of the Weinberg angle.
Both the expressions of $\delta x$ as function of the Wilson coefficients and the contribution of $\delta x$ to the EW observables are given in Appendix C of Ref.~\cite{Allwicher:2023aql}.
We include the running of the Wilson coefficients from the matching scale $\Lambda_{\rm UV}\sim 5\,$TeV to the EW scale and build a $\chi^2_{\rm EW}$ function,
\begin{equation}
\chi^2 = \sum_{ij}[O_{i,\text{exp}}-O_{i,\text{th}}] (\sigma_{\rm exp}^{2}+\sigma_{\rm th}^{2})^{-1}_{ij}[O_{j,\text{exp}}-O_{j,\text{th}}].\label{eq:chi2Def}
\end{equation}
We also consider the projections for the $Z$-pole observables at the Future Circular $e^+e^-$ Collider (FCC-ee)~\cite{FCC:2018evy,FCC:2018byv}. We expect similar conclusions for other tera-$Z$ machines
as CEPC~\cite{CEPCStudyGroup:2018ghi}.
We build a $\chi^2_{\rm FCC}$ with the projections estimated in Ref.~\cite{deBlas:2022ofj} as explained in Section 5.3 of Ref.~\cite{Capdevila:2024gki}.

The EW fit is mostly dominated by the tree level matching, so it is instructive to write the expressions neglecting the running.
Notice that although our model is intrinsically non-universal, we can divide the NP contributions into a universal part and a part acting only on the third family fields $q_L^3$ and $\tau_R$, which couple differently to the massive vector bosons. The universal part\footnote{Here, by universal NP, we mean that the only SM fermion currents that NP states can couple to are proportional to those the $W_{\mu}^a$ and $B_{\mu}$ SM bosons couple to.\label{FN:Univ}}
 can be described by the oblique parameters $\hat S$, $\hat T$, $Y$ and $W$ as defined in Ref.~\cite{Barbieri:2004qk} by rewriting the universal contributions in the SILH basis, whose Wilson coefficients are easily relatable to the oblique parameters~\cite{Giudice:2007fh}. Their expressions in the Warsaw basis are given in \cref{sec:AppOblique}.
Thus, using the Wilson coefficients of our model given in \cref{sec:ApSMEFT}, we obtain
\begin{align}
\hat S=&\frac{v^2}{8f^2}\left[\sin^2 (\theta_L)+\frac{g_L^2}{g_Y^2}\sin^2 (\theta_X) \right],~~ \hat T=\frac{v^2}{4f^2}c^2_H s^2_H,\\
 W=&\frac{v^2}{4f^2}\sin^4 (\theta_L),~~~~~~~~~~~~~~~~~~~~~~~~ Y=\frac{v^2}{4f^2} \frac{g_L^2}{g_Y^2}\sin^4 (\theta_X),
\end{align}
which affect $\delta m_W$ and $\delta g^{Z\psi}_{ij}=\delta g^{Z}_{\psi}\delta_{ij}$ as follows
\begin{align}
\delta m_W=&\,\frac{1}{2}\frac{g_L^2}{g_L^2-g_Y^2}\left(\hat T - \frac{g_Y^2}{g_L^2} Y\right)
-\frac{g_Y^2}{g_L^2-g_Y^2}\left(\hat S-Y-\frac{1}{2}W\right),\label{eq:WmassST}\\
\delta g^{Z}_{\psi}=&\,\frac{T_{\psi}^3}{2}\left(\hat T -\frac{g_Y^2}{g_L^2} Y  -W\right)
+\frac{Q_{\psi}}{2}\frac{g_Y^2}{g_L^2-g_Y^2} \left(\hat T-\frac{g_Y^2}{g_L^2} Y - 2\hat S+2 Y+ W \right).\label{eq:deltaST}
\end{align}
Here, $\psi$ denotes all SM fermions the $Z$ can decay into except $\tau_R$ and $b_L$ which receive extra non-universal contributions
\begin{align}
\Delta \delta g^{Z}_{b_L}\equiv&\,\delta g^{Z}_{b_L}-\left.\delta g^{Z}_{b_L}\right|_{\rm Univ.}=\frac{v^2}{16f^2}\cos(2\theta_L),\\
\Delta \delta g^{Z}_{\tau_R}\equiv\,&\delta g^{Z}_{\tau_R}-\left.\delta g^{Z}_{\tau_R}\right|_{\rm Univ.}=-\frac{v^2}{16f^2}\cos(2\theta_X).
\end{align}
If we try to decrease the scale $f$ as much as possible to reduce the tuning of the model on the Higgs mass, LHC searches will limit the mixing angles $\theta_L$ and $\theta_X$ to be small, suppressing $W$ and $Y$ with respect to $\hat S$ and $\hat T$. The fit in the relevant region is then dominated by $\hat S$ and $\hat T$ together with $\delta g^{Z}_{b_L}$ and $\delta g^{Z}_{\tau_R}$. Interestingly, the contribution of the massive vector bosons to $\hat T$ vanishes: the contributions from ${\cal B}_0$ and ${\cal B}_1^{\pm}$ cancel out, and the one from ${\cal W}$ is zero. The only contribution comes from a possible non-vanishing mixing angle between the Higgses.
It is also worth noticing that $\hat S$ is strictly positive, which implies a strictly negative shift in the $W$ boson mass as seen in~\cref{eq:WmassST}, being this observable one of the most  constraining ones. Then, a VEV for $H_2$ giving a positive contribution to $\hat T$ can actually alleviate a bit this tension and decrease the scale $f$.

\subsection{Quark flavor-changing observables}

The breaking of universality of the LH quark doublets can trigger dangerous FCNC processes. The situation in the quark sector is similar to the models studying a deconstructed $SU(2)_L$~\cite{Davighi:2023xqn,Capdevila:2024gki}, where one sees that the most constraining FCNC processes are $B_s$ mixing and $B_s\to \mu\mu$. Other processes like $B_d$ mixing or $K$ mixing are similar in magnitude but weaker.
How the breaking of universality affects FCNC processes depends on the alignment between the mass eigenstates of the quarks and interaction eigenstates to which NP couples. Let $(q_L^3)_I$ and $(q_L^{1,2})_I$ be the states charged under $SU(2)^{q_3}_L$ and $SU(2)_{L}^{\prime}$ respectively, and $(q_L^3)_d$ the state containing the mass eigenstate of the LH bottom quark. We can relate both states as
\begin{equation}
(q_L^3)_d =(q_L^3)_I-\epsilon_b \sum_{i=1,2}
(V^i_q)^* (q^i_L)_I ,
\end{equation}
where the parameter $\epsilon_b$ describes this alignment and $V_q^i$ is the spurion which breaks the flavor symmetry $U(2)_q$
and generates the CKM mixing elements. If we choose $(q_L^{1,2})_I$ aligned with the down-mass eigenstates (we have this freedom because our NP respects $U(2)_q$), then $V_q^i=(V_{td}^*,V_{ts}^*)$. For the explicit expression of the quark rotation matrices, see Eqs.~(11)--(15) of~\cite{Capdevila:2024gki}.
If $\epsilon_b=0\,(1)$, our third family quark doublet in the interaction basis is fully aligned to the mass eigenstate of the bottom (top). In the first case, FCNCs in the down sector vanish while in the second one are maximal.

Notice that our NP also affects the experimental determination of $V_{cb}$ from exclusive $B\to D^{(*)} \,\ell \nu$ and inclusive $B\to X_c \,\ell \nu$ decays:
\begin{equation}
V^{\rm Exp}_{cb}\equiv \frac{[C^{V,LL}_{\nu e d u}]^*_{ii32}}{[C^{V,LL}_{\nu e}]_{1221}} = V_{cb}
\left(1+\delta V_{cb}\right),~{\rm with}~\delta V_{cb}=(\epsilon_b-1)\frac{v^2}{8f^2},\label{eq:VcbExp}
\end{equation}
where $i=1,2$, $C^{V,LL}_{\nu e d u}$, $C^{V,LL}_{\nu e}$ are the Wilson coefficients of the LEFT operators
\begin{align}
O^{V,LL}_{\nu e d u} =(\bar \nu_L \gamma_{\mu} e_L)(\bar d_L \gamma^{\mu} u_L),~
O^{V,LL}_{\nu e} =(\bar \nu_L \gamma_{\mu} \nu_L)(\bar e_L \gamma^{\mu} e_L),
\end{align}
and $V_{cb}$ corresponds to the actual rotation element between the up and down basis. For values of $f> 1\,$TeV, $\delta V_{cb}$ is below $1\%$, smaller than the current error of $V_{cb}$ ($\sim 1\%-2\%$)~\cite{Bona:2024bue,ValeSilva:2024jml}, so this effect can be safely neglected.

Regarding $B_s$ mixing, in the low-energy effective field theory (LEFT) we can write the operator
\begin{equation}
\mathcal{L}=-C_{B_s} (V_{tb}V_{ts}^*)^2 (\bar s_L \gamma_{\mu} b_L)^2,
\end{equation}
which, at the EW scale receives the NP contributions from our model
\begin{equation}
C^{\rm NP}_{B_s}=\eta \frac{\epsilon_b^2}{8f^2},\label{eq:Csb}
\end{equation}
where $\eta \approx 0.84$ is the running from the matching scale $\sim 5\,$TeV to the EW scale~\cite{Fuentes-Martin:2020zaz}. Its contribution to $\Delta m_{B_{s}}$ is
\begin{equation}
\frac{\Delta m_{B_{s}}}{\Delta m_{B_{s}}^{\rm SM}}=\left| 1+
\frac{C^{\rm NP}_{B_{s}}}{C^{\rm SM}_{B_{d,s}}}\right|,
\end{equation}
where 
$
C_{B_{s}}^{\rm SM}=\frac{g_L^2}{32 \pi^2 v^2 } S_0,
$
with $S_0 \approx 2.49$~\cite{Buras:1998raa,Buchalla:1995vs}.
We then take~\cite{LHCb:2021moh,DiLuzio:2019jyq} 
\begin{equation}
\Delta m_{B_{s}}^{\rm Exp}=(17.7656\pm 0.0057)\,{\rm ps}^{-1},~~\Delta m_{B_{s}}^{\rm SM}=18.4^{+0.7}_{-1.2}\,{\rm ps}^{-1}.
\end{equation}
The other main observable, $B_s\to \mu \mu$, is affected by the LEFT operator,
\begin{equation}
\mathcal{L}\supset \frac{2}{v^2}V_{ts}^*V_{tb}\frac{\alpha_{\rm EM}}{4\pi}C_{10} (\bar s_L \gamma_{\mu} b_L) (\bar \mu \gamma^{\mu} \gamma^5 \mu),\label{eq:Lagbsll}
\end{equation}
receiving NP contributions in our model,
\begin{equation}
C^{\rm NP}_{10}=-\frac{\pi \epsilon_b}{\alpha_{\rm EM}}\frac{v^2}{8f^2}.\label{eq:C10}
\end{equation}
Its contribution to the observable is
\begin{equation}
\frac{{\cal B}(B_s\to \mu^+ \mu^-)}{{\cal B}(B_s\to \mu^+ \mu^-)_{\rm SM}}= \left|1+\frac{C_{10}^{{\rm NP}}}{C_{10}^{{\rm SM}}}\right|^2,\label{eq:Bsmumu}
\end{equation}
where $C_{10}^{\rm SM}=-4.19$~\cite{Isidori:2023unk} and~\cite{Neshatpour:2022pvg,Altmannshofer:2021qrr}
\begin{equation}
{\cal B}(B_s\to \mu^+ \mu^-)_{\rm Exp}=\left( 3.52^{+0.32}_{-0.30} \right) \times 10^{-9},~
{\cal B}(B_s\to \mu^+ \mu^-)_{\rm SM}=(3.67\pm 0.15)\times 10^{-9}.
\end{equation}
We can now build a $\chi^2_{bs}$ with these two observables following \cref{eq:chi2Def}.
Remarkably, the impact of NP on both observables only depends on $f$ and the degree of down-alignment $\epsilon_b$: the mixing angle $\theta_L$ cancels out in \cref{eq:Csb,eq:C10}.
Combining both observables, we obtain the $95\%$\,C.L. limits $f>2.0\,$TeV for $\epsilon_b=0.3$ (corresponding to a very mild down-alignment), $f>2.8\,$TeV for $\epsilon_b=0.5$ (intermediate alignment), and $f>4.8\,$TeV for $\epsilon_b=1$ (full up-alignment).

\subsection{Lepton flavor violation processes}

In analogy to the quark sector, the breaking of universality in the lepton sector is expected to imply lepton flavor violation (LFV) processes. In this case, NP couples non-universally to RH leptons but universally to LH leptons. It is interesting to note that this makes our model to avoid the stringent bounds from LFU tests in $\tau$ decays, which are mostly sensitive to NP non-universally coupled to LH leptons.

The light-Yukawa couplings written in \cref{eq:LightYuk} also induce terms like
\begin{equation}
\mathcal{L} \supset y_{3\mu}
{\bar\ell}_L^{\,3} H \mu_R + y_{3e} \bar \ell_L^{\,3} H e_R+{\rm h.c.},
\end{equation}
where we expect $y_{3\mu}\sim y_{\mu}$ and $y_{3e}\sim y_{e}$. This generates chirally suppressed mixing angles between the interaction eigenstates and the mass eigenstates of the RH leptons, $\theta_{23}\sim m_{\mu}/m_{\tau}$ and $\theta_{13}\sim m_{e}/m_{\tau}$, which, in turn, generates LFV vertices of the ${\cal B}_0$ vector boson in the mass basis of the RH fermions:
\begin{equation}
\mathcal{L}\supsetsim g_Y \csc(2\theta_X)\, \mathcal{B}_{0\mu} \left(
\frac{m_{\mu}}{m_{\tau}}
\bar \tau_R  \gamma^{\mu}  \mu_R
+
\frac{m_{e}}{m_{\tau}}
\bar \tau_R  \gamma^{\mu}  e_R+
\frac{m_{\mu}m_e}{m_{\tau}^2} \bar \mu_R \gamma^{\mu}  e_R
+{\rm h.c.}\right).\label{eq:LFVg}
\end{equation}
When ${\cal B}_{0}$ is integrated out, these couplings generate operators in the LEFT:
\begin{align}
\mathcal{L}\supset  \sum_{\substack{\psi=e,u,d\\K=L,R}}
C_{e\psi}^{V,RK} (\bar e_R\gamma_{\mu} e_R)(\bar \psi_K\gamma^{\mu} \psi_K) ,
\end{align}
with relevant LFV Wilson coefficients,
\begin{align}
[C_{e\psi}^{V,RK}]_{ijkk}=&-\frac{m_{\ell_i} m_{\ell_j}}{4 m_{\tau}^2 f^2} \left(Y_{\psi_K}+Q_\psi(s_W^2-1)\cos(2\theta_X)\right),\label{eq:LFVOp}
\end{align}
with $\psi=e,u,d$, $K=L,R$ and $j>i\geq k=1,2$. 
These Wilson coefficients generate LFV three-body leptonic decays with branching fractions~\cite{Crivellin:2013hpa}:
\begin{align}
{\cal B}(\ell_{k}\to \ell_j \bar\ell_i \ell_i) &= \frac{M_{\ell_k}^5}{1536 \pi^3 \Gamma_{\ell_k}(1+\delta_{ij})}\bigg(\left|[C_{ee}^{V,RL}]_{ijkk}\right|^2
+\left|[C_{ee}^{V,RR}]_{ijkk}\right|^2\bigg),
\end{align}
and $\mu\to e$ conversion~\cite{Kuno:1999jp}, for which the branching fractions can be computed from the formulas given in Ref.~\cite{Ardu:2024bua}.
Among these, the main experimental bounds in our model come from $\tau$ LFV decays
~\cite{HFLAV:2019otj}:
\begin{align}
\mathcal{B}(\tau^-\to \mu^-\mu^+\mu^-)<\,1.1\times 10^{-8},~~
\mathcal{B}(\tau^-\to \mu^-e^+e^-)<\,1.1\times 10^{-8},\label{eq:taumu2e}
\end{align}
and $\mu \to e$ conversion in the SINDRUM~II experiment with gold nuclei~\cite{SINDRUMII:2006dvw}:
\begin{equation}
{\cal B}(\mu \,{\rm Au}\to e\,{\rm Au}) < 7 \times 10^{-13},
\label{eq:muAu2eAu}
\end{equation}
all of them at $90\%$C.L. We construct an associated $\chi^2_{\rm LFV}$ function as
\begin{equation}
\chi^2_{\rm LFV}=F^{-1}_{\chi^2}(1;0.9)\sum_i \frac{O_{i,{\rm th}}^2}{O_{i,90\%}^2},
\end{equation}
where $F^{-1}_{\chi^2}(n;x)$ is the inverse cumulative distribution function of a $\chi^2$ with $n$ degrees of freedom (d.o.f) and $O_{90\%}$ is the experimental limit at $90\%$\,C.L.

Future measurements plan to increase substantially the sensitivity to these processes.
The Mu3e experiment~\cite{Hesketh:2022wgw} aims to reach $\mathcal{B}(\mu^+\to e^+e^-e^+)<10^{-16}$ at $90\%$ C.L. in 2028 in case of lack of signal (the current limit is $10^{-12}$).
Belle II expects to improve the bounds on $\tau$ LFV decays to~\cite{Belle-II:2018jsg}
\begin{align}
\mathcal{B}(\tau^- \to \mu^-\mu^+\mu^-)<4.6 \times 10^{-10},~~
\mathcal{B}(\tau^- \to \mu^- e^+e^-)<3.0 \times 10^{-10},
\end{align}
at $90\,\%$ C.L. for an integrated luminosity of $50$ ab$^{-1}$.
Regarding $\mu\to e$ conversion, the COMET~\cite{COMET:2009qeh} and Mu2e~\cite{Mu2e:2014fns} experiments aim to reach a sensitivity of $10^{-16}-10^{-17}$ in Aluminium. Conservatively, we take for the projection $\mathcal{B}(\mu\,{\rm Al} \to e \, {\rm Al})<10^{-16}$ at $90\%$~C.L.

\subsection{Results and projections}

In this section we show the exclusion limits for every set of observables defined above, constructing the corresponding $\chi^2$.
The $95\%$\,C.L. exclusion limits are then defined by $\chi^2-\chi^2_{\rm min}=F^{-1}_{\chi^2}(n_{\rm dof};0.95)$. We take for $n_{\rm dof}$ the number of linearly independent functions of the parameters that the different observables in the considered likelihood depend on. An example of this rule is that for limits set by only one observable, we use 1 d.o.f. because we can consider the single observable as the only independent function of the parameters.

\begin{figure*}[t]
\begin{tabular}{cc}
\includegraphics[width=0.48\textwidth]{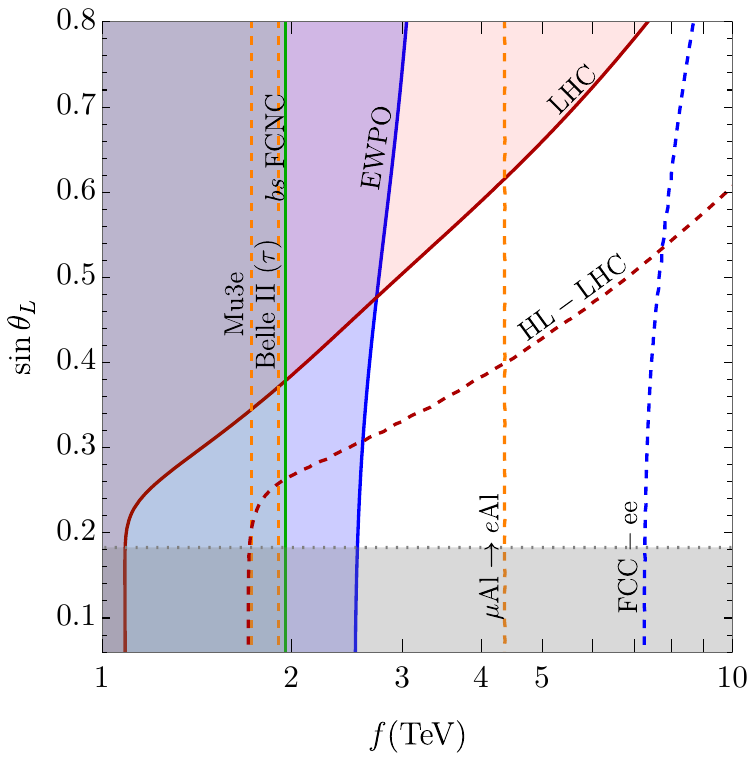} &
\includegraphics[width=0.48\textwidth]{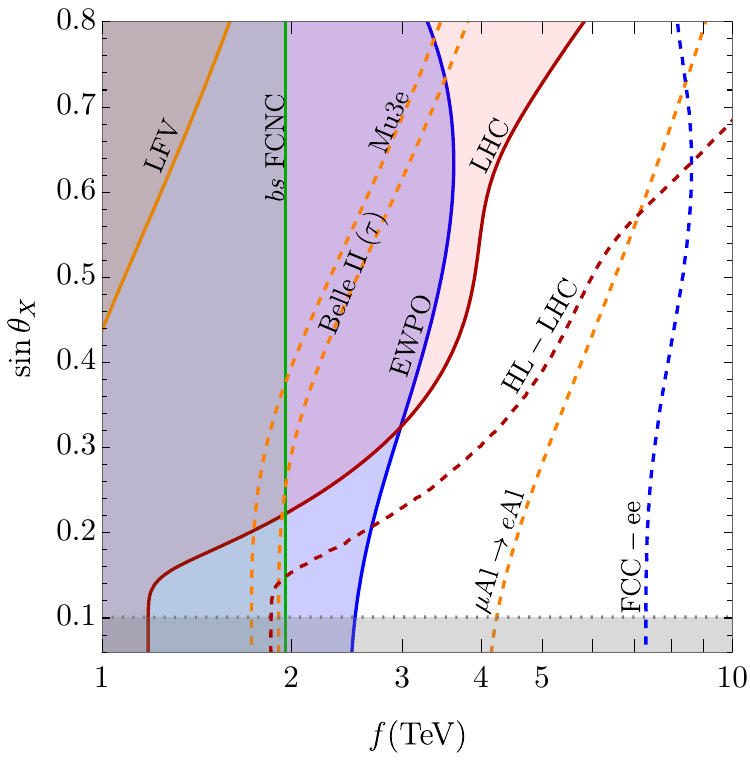} 
\end{tabular}
\caption{Exclusion limits fixing $\langle H_2 \rangle=0$, and $\sin(\theta_X)=g_Y/g_{R2}=g_Y/2.5$ (left) and $\sin(\theta_L)=g_L/g_{L2}=g_L/2.5$ (right). Colored regions are excluded at the $95\%\,$C.L. and dashed lines show projections of future measurements for the $95\%\,$C.L. limits. 
For $bs$ FCNC we assume $\epsilon_b=0.3$ (very mild down-alignment).
The gray areas depict regions where $g^2/(4\pi)>1$ for some of the fundamental gauge couplings, so the model may enter in a non-perturbative regime. We take 3 d.o.f. for EWPO, FCC-ee, LHC and HL-LHC searches, 2 for current LFV limits, and 1 for the others.}
\label{fig:Fixed}
\end{figure*}

The phenomenology of our model is mostly described by 4 parameters: the Higgs decay constant $f$, the two mixing angles $\theta_L$ and $\theta_X$ of \cref{eq:thetaLX}, and the mixing angle of the two Higgses $|s_H|=|\langle H_2 \rangle|/v$. This last one, contributing to $\hat T$, could vanish depending on the extended sector that generates the scalar potential as discussed in~\cref{sec:AppPot,sec:EHC}. In the first part of the analysis we fix $s_H= 0$, or equivalently, $\langle H_2 \rangle = 0$ as a benchmark.
\cref{fig:Fixed} shows the $95\,\%$ C.L. excluded regions for the different likelihoods for fixed values of $\sin(\theta_X)=g_Y/g_{R2}=g_Y/2.5$ (left) and $\sin(\theta_L)=g_L/g_{L2}=g_L/2.5$ (right) as function of the other mixing angle and the decay constant of the Higgs, $f$.
The main constraints come from LHC searches for larger values of the mixing angles $\theta_{L,X}$ and EWPO for smaller values. 
We can see how values of $f=(2.5-3)\,$TeV are phenomenologically acceptable. 
FCNC in the quark sector are subdominant assuming a very mild down-alignment of $30\%$. In the lepton sector, FCNC processes are also well below other bounds. This situation however can change with the up-coming experiments COMET and Mu2e testing $\mu \to e$ conversion in Aluminium, with a sensibility capable to test scales of the Higgs decay constant of $4-9$\,TeV.
HL-LHC will improve the bounds for medium and large mixing angles $\theta_{L,X}$, but for small mixing angles, EWPO from LEP will still dominate. The $Z$-pole measurements of FCC-ee could however increase significantly the bound on $f$ to $\sim (7-9)$\,TeV.

\begin{figure*}[t]
\begin{tabular}{cc}
\includegraphics[width=0.48\textwidth]{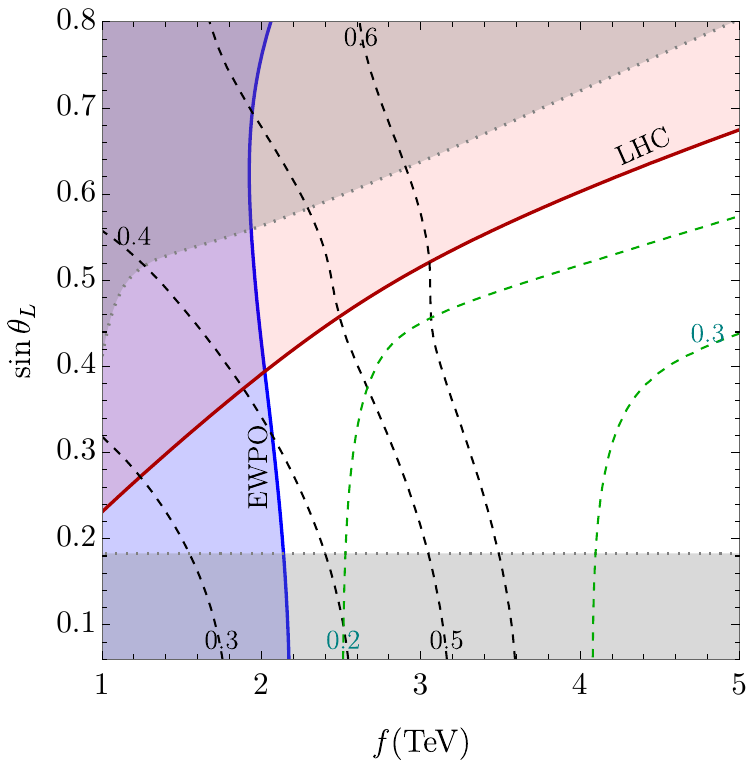} &
\includegraphics[width=0.48\textwidth]{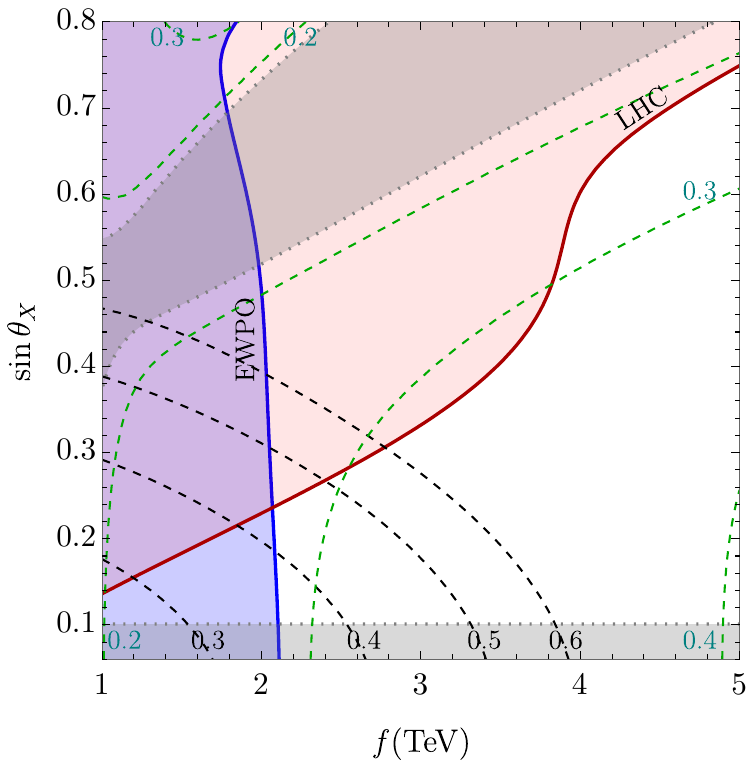} 
\end{tabular}
\caption{Profiled exclusion limits as function of $f$ and $\theta_{L(X)}$ on the left (right). Colored regions are excluded at the $95\%\,$C.L. for 4 d.o.f.
Dashed lines show contours of the non-shown parameters whose values minimize the likelihood, $|s_H|$ in black, and the mixing angle $\theta_{X(L)}$ on the left (right) in green.
The gray areas depict regions where $g^2/(4\pi)>1$ for some of the fundamental gauge couplings, so the model may enter in a non-perturbative regime.}
\label{fig:Prof}
\end{figure*}

We now consider the more general case allowing for $\langle H_2 \rangle \neq 0$. This contributes positively to $\hat T$, but some amount of $\hat T$ can improve EWPO because it partially cancels the contribution of a positive $\hat S$ to the $W$ boson mass and some $Z$ vertex modifications~(see \cref{eq:WmassST,eq:deltaST}). To test the most optimistic situation, we combine the $\chi^2$ of LHC searches and EWPO, and show the $95\%\,$C.L. exclusion regions in \cref{fig:Prof} as function of one of the mixing angles $\theta_{L,X}$ and $f$ after profiling over the other mixing angle and $s_H$. This is, for every point in these figures, we find the minimum of $\chi^2_{\rm EW}+\chi^2_{\rm LHC}$ fixing thereby the other two parameters. We can see that actually, $O(1)$ mixing angles $s_H$ decrease the bound. In particular, limits on $f$ can go down to 
$\sim 2\,$TeV for $|s_H| \approx 0.3-0.4$. We have further checked that even a maximal mixing angle, $|s_H|=\sqrt{2}/2$, reduces the bound with respect to $s_H=0$.\footnote{The alternative model with a gauged $U(1)_{R2}$ instead of the full $SU(2)_{R2}$ does not have the ${\cal B}_1$ vector field, but the charged pNGB $\Delta^{\pm}$. The oblique parameter $\hat T$ does not receive the ${\cal B}_1$ contribution, getting positively shifted by $\Delta \hat T=v^2/16f^2$. This reduces the limit of $f$ in this model to $\sim 2.3\,$TeV in the case $\langle H_2 \rangle = 0$.\label{foot:U1R}}

\section{Conclusions}
\label{sec:Conclusions}

Composite Higgs theories are one of the most promising theories for NP to address the Higgs hierarchy problem. However, incorporating flavor is still challenging. Although these models present mechanisms to produce flavor hierarchies, like anarchic partial compositeness~\cite{Grossman:1999ra,Gherghetta:2000qt}, or dynamical scales~\cite{Panico:2016ull,Fuentes-Martin:2022xnb}, an explanation of why the specific patterns of masses and mixings of the SM emerge typically remains unanswered: Why the different fermions are arranged into families of increasing masses? Why CKM matrix is close to the identity while the PMNS matrix is anarchic?

We have presented a composite Higgs model that intrinsically addresses some of this flavor patterns by its own construction due to its extended non-universal gauge symmetry. Leading Yukawa couplings can only be written for the third family while light-family Yukawa couplings require higher dimensional operators generated at higher scales. Furthermore, due to the EW charges of the fermionic degrees of freedom of the new composite sector, the model has the adequate accidental flavor symmetries to address the hierarchies of CKM and PMNS matrices simultaneously.

The new composite sector we propose is minimal: it is simply a copy of QCD, confining slightly above the TeV scale, and with four flavors of hyper-quarks. Still, it is capable to address three independent but crucial tasks in our model:
\begin{itemize}
\item It provides a radiatively stable sector that breaks the extended non-universal gauge symmetry to the SM one. This is already advantageous with respect to other models implementing flavor deconstruction that use scalars to break the UV gauge symmetry, introducing new hierarchy problems.
\item It provides a mechanism to arrange the SM fields in representations that look anomalous, but that are not when the hyper-quarks are taken into account . This allows to generate the accidental flavor symmetry $U(2)_q\times U(2)_e$, crucial to address minimally the hierarchies of masses and CKM mixing elements between third and light families, but keeping  the PMNS matrix anarchic.
\item It gives a composite Higgs that triggers the EW symmetry breaking of the SM, solving the large hierarchy problem of the Higgs. This links the breaking scale of the UV gauge symmetry and the Higgs decay constant $f$ to be the same.
\end{itemize}

The phenomenology is mostly driven by the presence of new massive vector boson with EW charges above the TeV. Our model is currently mostly constrained by EW precision data from LEP, which is affected mainly by these new massive vector bosons, and by their LHC searches in $pp\to $\,leptons. The scale $f$ has a minimal experimental bound of $(2-3)$\,TeV, depending on the details of the model, requiring a fine tuning of $10^{-2}-10^{-3}$ to get an EW mass for the SM Higgs boson.
Perhaps this tuning could be relaxed by adding vector-like quarks implementing a collective breaking also for the top-quark Yukawa.
Other subdominant bounds come from FCNC in the quark and the lepton sectors. Interestingly, we expect an amazing improvement in the sensitivity to LFV processes in the coming years, particularly in $\mu \to e$ conversion, which could test scales for $f$ of $4 -9$\,TeV, even assuming the most pessimistic projection. We can conclude that this observable will become a prime test for NP at the TeV scale breaking the universality between taus and light leptons if chirally suppressed rotations between the interaction and mass bases of charge leptons are also present (see also Ref.~\cite{Capdevila:2024gki}). Other future projections, like searches in HL-LHC reduce the available parameter space, but do not change qualitatively the current picture. Tera-$Z$ machines like FCC-ee or CEPC could however test scales of $f\sim 8$\,TeV.

Our model is a UV completion of the SM at the TeV scale, but we should see it already as an effective theory. We expect higher scale NP addressing different aspects of the model that we have not carefully studied in this paper and we leave for future work.
One of them is the extended hyper-color sector producing the necessary four-fermion operators for the top Yukawa coupling and the Higgs potential that should appear at $\Lambda_{\rm HC}\sim 20\,$TeV. As discussed in \cref{sec:EHC}, scalar and vectors can make the job, but their existence would raise new questions, like their radiative stability or their origin.
Maybe a specific realization of the extended sector establishes relations among the couplings contributing to alleviate the naive little tuning commented above.
We have also left unspecified the physics that UV-completes the six-fermion operators necessary to write light-family Yukawa couplings, but briefly commented on possibilities in \cref{sec:EHC} with hierarchically heavier vector-like fermions or scalars with masses $10^{2-3}$\,TeV. However, the origin of hierarchies between first and second families remain unaddressed. Perhaps the same mechanism could be copied, and we could UV-complete the factors charging the light families so $SU(2)^{\prime}_L\times U(1)_X \leftarrow SU(2)^{(2)}_L\times  SU(2)^{(1)}_L \times SU(2)_{R}^{\ell_2} \times U(1)^{\prime}_X $ at some higher scale $\sim 100\,$TeV, similarly to the proposals in Refs.~\cite{Bordone:2017bld,FernandezNavarro:2023rhv,Davighi:2023xqn}.

Furthermore, we have not explored possible connections of our model with other open puzzles. For instance, besides the pNGB at the scale $f$, the composite sector will also generate resonances like hyper-baryons at the scale $\Lambda_{\rm HC}\sim 20\,$TeV, with a lightest hyper-baryon that could be neutral under the SM gauge group. If there is conservation of hyper-baryon number, one could wonder if this could be a constituent of dark matter (DM)~\cite{Nussinov:1985xr,Chivukula:1989qb,Banks:2010eh,Ma:2015gra}. However, the latest bounds from DM direct searches~\cite{LZ:2022lsv} put a limit of around $100\,$TeV for singlet fermions with a magnetic dipole moment similar to the neutron (in units of their respective inverse masses)~\cite{Eby:2023wem}.\footnote{Possible electric dipole moments give limits $2$ orders of magnitude stronger in the hyper-baryon mass~\cite{Antipin:2014qva}. However, they can be avoided imposing CP conservation.}
If this is the case, our stable hyper-baryon could only be a small fraction of the total DM.

\acknowledgments

We would like to thank Mikael Chala, Joe Davighi, Xabier Marcano, Javier Fuentes-Mart\'in, Gino Isidori, Andres D. Perez, Manuel Pérez-Victoria, Stefan Pokorski, Rosa M. Sandá Seoane, Nud{\v z}eim Selimovi{\'c}, Javi Serra and Peter Stangl for useful discussions. We further thank Javier Fuentes-Mart\'in and Javi Serra for useful comments on the manuscript.
This work has been supported by the grant CSIC-20223AT023. We also acknowledge the support of the Spanish Agencia Estatal de Investigacion through the grant “IFT Centro de Excelencia Severo Ochoa CEX2020-001007-S”.

\appendix

\crefalias{section}{appendix}

\section{Semisimple UV completion}
\label{sec:SemisimpleUV}

Our model can be completed in a far UV above the TeV scale to the semisimple gauge group $SU(2)^4\times SU(4)\times SU(N_{\rm HC}+1)$ as shown in \cref{tab:SemisimpleUVCompl}. The group $SU(3)_c\times \mathcal{G}_{2^4 1}$ (see \cref{eq:G241}) is a subgroup: $SU(3)_c\times U(1)^{\prime}_{B-L} \subset SU(4)$ \`a la Pati-Salam~\cite{Pati:1974yy}, and $SU(N_{\rm HC})\times U(1)_{\rm HC} \subset SU(N_{\rm HC}+1)$, with $U(1)_{B-L}=[U(1)^{\prime}_{B-L}\times U(1)_{\rm HC}]_{\rm diag}$. 
Light-family leptons are unified with light-family quarks into a fundamental of $SU(4)_{\rm PS}$, while third-family leptons are unified with the hyper-quarks into a fundamental of $SU(N_{\rm HC}+1)$. This difference explains their anomalous quantum numbers under $\mathcal{G}_{2^4 1}$. The embedding requires the introduction of the leptons $L_{L,R}^{(L,R)}$ which will get a mass after the breaking of the far-UV gauge symmetry,
$
\mathcal{L}\supset M_L\bar L_L^{(L)}L_R^{(L)}+M_R\bar L_L^{(R)}L_R^{(R)}.
$
They become vector-like leptons of our model with a mass at the far UV breaking scale. Similarly to the Pati-Salam completion of the SM, this UV completion makes transparent the anomaly cancellation of our model. Since we only have fundamental representations, we only have to check that $SU(4)_{\rm PS}$ and $SU(N_{\rm HC}+1)$ charge the same number of LH and RH fields, and that the $SU(2)$ factors charge an even number of fields.

\begin{table}[t]
\renewcommand{\arraystretch}{1.2}
\begin{center}
\begin{tabular}{|c||c|c||c|c||c|c|}
\hline
Field  & $SU(2)_{L1}$ & $SU(2)_{R1}$ & $SU(2)_{L2}$ &  $SU(2)_{R2}$ &  $SU(4)_{\rm PS}$ & $SU(N_{\rm HC}+1)$\\
\hline
\hline
$(q^3_L,L^{(L)}_L)$  & $\mathbf{1}$ & $\mathbf{1}$ & $\mathbf{2}$ & $\mathbf{1}$ & $\mathbf{4}$ & $\mathbf{1}$\\
\hline
$(q^{1,2}_L,\ell_{L}^{1,2})$  & $\mathbf{2}$ & $\mathbf{1}$ & $\mathbf{1}$ & $\mathbf{1}$ & $\mathbf{4}$ & $\mathbf{1}$\\
\hline
$(q^3_R,L^{(R)}_R)$ & $\mathbf{1}$ & $\mathbf{2}$ & $\mathbf{1}$ & $\mathbf{1}$ & $\mathbf{4}$ & $\mathbf{1}$  \\
\hline
$(q_R^{1,2},\ell_R^{1,2})$ & $\mathbf{1}$ & $\mathbf{2}$ & $\mathbf{1}$ & $\mathbf{1}$ & $\mathbf{4}$ & $\mathbf{1}$  \\
\hline
\hline
$(\zeta_L^{(L)},\ell^3_L)$ & $\mathbf{2}$ & $\mathbf{1}$ & $\mathbf{1}$ & $\mathbf{1}$ &  $\mathbf{1}$ &  $\square$ \\
\hline
$(\zeta_L^{(R)},L^{(R)}_L)$  & $\mathbf{1}$ & $\mathbf{2}$ & $\mathbf{1}$  & $\mathbf{1}$ & $\mathbf{1}$ & $\square$ \\
\hline
$(\zeta_R^{(L)},L^{(L)}_R)$ & $\mathbf{1}$  & $\mathbf{1}$ & $\mathbf{2}$ & $\mathbf{1}$ & $\mathbf{1}$ & $\square$  \\
\hline
$(\zeta_R^{(R)},\ell^3_R)$ & $\mathbf{1}$ & $\mathbf{1}$ & $\mathbf{1}$ & $\mathbf{2}$ & $\mathbf{1}$  & $\square$ \\
\hline
\hline
$N_L$ & $\mathbf{1}$ & $\mathbf{1}$ & $\mathbf{1}$ & $\mathbf{1}$ & $\mathbf{1}$  & $\mathbf{1}$ \\
\hline
\end{tabular}
\end{center}
\caption{Semisimple UV completion for our model given in \cref{tab:fieldcontent}. We use $\square$ to denote the fundamental representation.}
\label{tab:SemisimpleUVCompl}
\end{table}

\section{An alternative model}
\label{sec:AppModelII}

There is an alternative model implementing the accidental $U(2)_q\times U(2)_e$ flavor symmetry, satisfying conditions (1)-(4) of \cref{sec:AFD}, but not (5). It is shown in \cref{tab:fieldcontentII} (top) arranging the fields in representations of the global group $\mathcal{G}_{2^41}$. All perturbative gauge anomalies cancel, and global anomalies cancel if $N_{\rm HC}$ is even. Similarly to the model of the paper, this also implies that resonances of the hyper-sector have integer electric charge, $Q=T_L^3+T_R^3+B-L$: in this case, hyper-baryons (hyper-mesons) have $B-L=1(0)$ and $T_L^3+T_R^3$ is integer.

We also show a possible UV completion into a semisimple group in \cref{tab:fieldcontentII} (bottom), analogous to what is done in \cref{sec:SemisimpleUV} for the model studied in the paper. In a far UV, $SU(N_{\rm HC}+2)$ breaks into $SU(N_{\rm HC})\times U(1)_{\rm HC}$, where $\zeta$ and the fields $\{\ell,L^{(\prime)}\}$ in those multiplets have charges $1/N_{\rm HC}$ and $-1/2$ respectively under $U(1)_{\rm HC}$. Also, $SU(4)_{\rm PS}$ breaks into $SU(3)_c\times U(1)^{\prime}_{B-L}$ like in Pati-Salam unification, and $U(1)_{B-L}=[U(1)_{\rm HC}\times U(1)^{\prime}_{B-L}]_{\rm diag}$.
Finally, the fields $L^{1,2}_{L,R}$ and  $L^{\prime\,1,2}_{L,R}$ can get a Dirac mass in the far UV at the scale of the breaking $U(1)_{\rm HC}\times U(1)^{\prime}_{B-L}\to U(1)_{B-L}$ and become vector-like leptons.

\begin{table}
\renewcommand{\arraystretch}{1.2}
\begin{center}
\begin{tabular}{|c||c|c||c|c||c|c||c|}
\hline
Field  & $SU(2)_{L1}$ & $SU(2)_{R1}$ & $SU(2)_{L2}$ & $SU(2)_{R2}$ & $U(1)_{B-L}$ & $SU(3)_c$ & $SU(N_{\rm HC})$\\
\hline
\hline
$q^3_L$ & $\mathbf{2}$ & $\mathbf{1}$ & $\mathbf{1}$ & $\mathbf{1}$ & $1/6$ & $\mathbf{3}$  & $\mathbf{1}$\\
\hline
$q^{1,2}_L$  & $\mathbf{1}$ & $\mathbf{1}$ & $\mathbf{2}$ & $\mathbf{1}$ & $1/6$ & $\mathbf{3}$ & $\mathbf{1}$\\
\hline
$q^{1,2,3}_R$  & $\mathbf{1}$ & $\mathbf{2}$ & $\mathbf{1}$ & $\mathbf{1}$ & $1/6$ & $\mathbf{3}$ & $\mathbf{1}$ \\
\hline
$\ell^{1,2,3}_L$ & $\mathbf{2}$ & $\mathbf{1}$ & $\mathbf{1}$ & $\mathbf{1}$ & $-1/2$ & $\mathbf{1}$ & $\mathbf{1}$ \\
\hline
$\ell^3_R$ & $\mathbf{1}$ & $\mathbf{2}$ & $\mathbf{1}$ & $\mathbf{1}$ & $-1/2$ & $\mathbf{1}$ & $\mathbf{1}$\\
\hline
$\ell^{1,2}_R$ & $\mathbf{1}$ & $\mathbf{1}$ & $\mathbf{1}$ & $\mathbf{2}$ & $-1/2$ & $\mathbf{1}$ & $\mathbf{1}$ \\
\hline
\hline
$\zeta_L^{(L)}$  & $\mathbf{2}$ & $\mathbf{1}$ & $\mathbf{1}$ & $\mathbf{1}$ & $1/N_{\rm HC}$ & $\mathbf{1}$ & $\square$\\
\hline
$\zeta_L^{(R)}$  & $\mathbf{1}$ & $\mathbf{2}$ & $\mathbf{1}$ & $\mathbf{1}$ & $1/N_{\rm HC}$ & $\mathbf{1}$ & $\square$ \\
\hline
$\zeta_R^{(L)}$  & $\mathbf{1}$ & $\mathbf{1}$ & $\mathbf{2}$ & $\mathbf{1}$ & $1/N_{\rm HC}$ & $\mathbf{1}$ & $\square$\\
\hline
$\zeta_R^{(R)}$  & $\mathbf{1}$ & $\mathbf{1}$ & $\mathbf{1}$ & $\mathbf{2}$ & $1/N_{\rm HC}$ & $\mathbf{1}$ & $\square$\\
\hline
\end{tabular}
\end{center}
\begin{center}
\begin{tabular}{|c||c|c||c|c||c||c|}
\hline
Field  & $SU(2)_{L1}$ & $SU(2)_{R1}$ & $SU(2)_{L2}$ &  $SU(2)_{R2}$ &  $SU(4)_{\rm PS}$ & ${SU{(N_{\rm HC}+2)}}$\\
\hline
\hline
$(q^3_L,\ell^3_L)$  & $\mathbf{2}$ & $\mathbf{1}$ & $\mathbf{1}$ & $\mathbf{1}$ & $\mathbf{4}$ & $\mathbf{1}$\\
\hline
$(q^{1,2}_L,L_{L}^{1,2})$  & $\mathbf{1}$ & $\mathbf{1}$ & $\mathbf{2}$ & $\mathbf{1}$ & $\mathbf{4}$ & $\mathbf{1}$\\
\hline
$(q^3_R,\ell^{3}_R)$ & $\mathbf{1}$ & $\mathbf{2}$ & $\mathbf{1}$ & $\mathbf{1}$ & $\mathbf{4}$ & $\mathbf{1}$  \\
\hline
$(q_R^{1,2},L^{\prime 1,2}_R)$ & $\mathbf{1}$ & $\mathbf{2}$ & $\mathbf{1}$ & $\mathbf{1}$ & $\mathbf{4}$ & $\mathbf{1}$  \\
\hline
\hline
$(\zeta_L^{(L)},\ell^1_L,\ell^2_L)$ & $\mathbf{2}$ & $\mathbf{1}$ & $\mathbf{1}$ & $\mathbf{1}$ &  $\mathbf{1}$ &  $\square$ \\
\hline
$(\zeta_L^{(R)},L^{\prime 1}_L,L^{\prime 2}_L)$  & $\mathbf{1}$ & $\mathbf{2}$ & $\mathbf{1}$  & $\mathbf{1}$ & $\mathbf{1}$ & $\square$ \\
\hline
$(\zeta_R^{(L)},L^{1}_R,L^{2}_R)$ & $\mathbf{1}$  & $\mathbf{1}$ & $\mathbf{2}$ & $\mathbf{1}$ & $\mathbf{1}$ & $\square$  \\
\hline
$(\zeta_R^{(R)},\ell^1_R,\ell^2_R)$ & $\mathbf{1}$ & $\mathbf{1}$ & $\mathbf{1}$ & $\mathbf{2}$ & $\mathbf{1}$  & $\square$ \\
\hline
\end{tabular}
\end{center}
\caption{Alternative model with an accidental $U(2)_q\times U(2)_e$ flavor symmetry. On the top we show the arrangement of the fields under the global group ${\cal G}_{2^41}$. On the bottom we show a possible semisimple UV completion. We use $\square$ to denote the fundamental representation.}
\label{tab:fieldcontentII}
\end{table}

In this model, gauge symmetry restricts the possible four-fermion operators generating Yukawa couplings to the pNGB Higgses to be
\begin{align}
\mathcal{L} & \supset \frac{1}{\Lambda_{u}^2}  \sum_{\substack{i=1,2,3 \\ j=1,2}}\left[\lambda_{u,ji}(\bar q^j_{L,\alpha} u_R^i) \delta^{\alpha \beta} (\bar \zeta_{L,1}^{(R)} \zeta_{R,\beta}^{(L)}) 
+ \lambda_{u,ji}(\bar q^j_{L,\alpha } u_R^i) \epsilon^{\alpha \beta} (\bar \zeta_{R,\beta}^{(L)} \zeta_{L,2}^{(R)})\right] \nonumber\\ 
&+ 
\frac{1}{\Lambda_{d}^2}
\sum_{\substack{i=1,2,3 \\ j=1,2}}\left[ \lambda_{d,ji}(\bar q^j_{L,\alpha } d_R^i) \delta^{\alpha \beta} (\bar \zeta_{L,2}^{(R)} \zeta_{R,\beta}^{(L)}) 
+ \lambda_{d,ji} (\bar q^j_{L,\alpha } d_R^i) \epsilon^{\alpha \beta} 
(\bar \zeta_{R,\beta}^{(L)} \zeta_{L,1}^{(R)})\right]  \nonumber\\
&+ 
\frac{1}{\Lambda_{e}^2}
\sum_{\substack{i=1,2,3 \\ j=1,2}} \left[ \lambda_{e,ij}(\bar \ell^i_{L,\alpha } e_R^j) \delta^{\alpha \beta} (\bar \zeta_{R,2}^{(R)} \zeta_{L,\beta}^{(L)}) 
+\lambda_{e,ij} (\bar \ell^i_{L,\alpha } e_R^j) \epsilon^{\alpha \beta} 
(\bar \zeta_{L,\beta}^{(L)} \zeta_{R,1}^{(R)})\right] 
+{\rm h.c.}\label{eq:ModelII4ferYuk}
\end{align}
They can only involve light families, $q_L^{1,2}$ and $\ell_R^{1,2}$, as it is illustrated in \cref{tab:Illustration2}, and
third-family Yukawa couplings to the pNGB Higgses cannot be generated with four-fermion operators.
In this model we can only generate sizable third-family Yukawa couplings by adding another scalar field bi-doublet of $SU(2)_{L1}\times SU(2)_{R1}$ playing the role of SM Higgs. This field could also be generated by extending the composite sector as in Ref.~\cite{Fuentes-Martin:2020bnh} with more hyper-quarks implementing additionally the breaking of a group $G \to H$, so $SU(2)_{L1}\times SU(2)_{R1} \subset H$. 

%%%%%
\begin{table}[t]
\renewcommand{\arraystretch}{1.2}
\begin{center}
\begin{tabular}{|c||c|c|}
\hline
 & Site 1 ($\zeta_L$) & Site 2 ($\zeta_R$)  \\
\hline
\hline
$SU(2)_L$  & $q_L^{3},~\ell_L^{1,2,3}$ & $q_L^{1,2}$  \\
\hline
$SU(2)_R$ & $q_R^{1,2,3},~\ell_R^{3}$ & $\ell_R^{1,2}$ \\
\hline
\end{tabular}
\end{center}
\caption{Arrangement of the SM fields on the sites 1 and 2 (corresponding to the factors that charge the hyper-quarks $\zeta_L$ and $\zeta_R$ respectively) for this alternative model. Since the operator interpolating the Higgs is $\mathcal{O}_H \sim \bar \zeta_L \zeta_R$, it can only couple fields in the diagonals of the table due to the deconstructed gauge symmetry.}
\label{tab:Illustration2}
\end{table}
%%%%%

\section{pNGB potential}
\label{sec:AppPot}

Expanding \cref{eq:V1}, we obtain the contributions to the pNGB masses
\begin{align}
\Delta m_S^2=&\,\frac{a_f f^2}{32\pi^2}\frac{\Lambda_{\rm HC}^4}{\Lambda_{N}^4}\,{\rm Re}(\lambda^*_N\lambda_N^{\prime}),\label{eq:mSL}\\
\Delta m_{H_1}^2=&\frac{a_f f^2}{64\pi^2}\left[
-N_c\frac{\Lambda_{\rm HC}^4}{\Lambda_t^4}|\lambda_t+\lambda_t^{\prime}|^2
-N_c\frac{\Lambda_{\rm HC}^4}{\Lambda_b^4}|\lambda_b-\lambda_b^{\prime}|^2
-\frac{2\Lambda_{\rm HC}^4}{\Lambda_{\tau}^4}|\lambda_{\tau}+\lambda_{\tau}^{\prime}|^2
+\frac{\Lambda_{\rm HC}^4}{\Lambda_N^4}|\lambda_N+\lambda_N^{\prime}|^2
\right],\label{eq:mH1L}\\
\Delta m_{H_2}^2=&\,\frac{a_f f^2}{64\pi^2}\left[
-N_c\frac{\Lambda_{\rm HC}^4}{\Lambda_t^4}|\lambda_t-\lambda_t^{\prime}|^2
-N_c\frac{\Lambda_{\rm HC}^4}{\Lambda_b^4}|\lambda_b+\lambda_b^{\prime}|^2
-\frac{2\Lambda_{\rm HC}^4}{\Lambda_{\tau}^4}|\lambda_{\tau}-\lambda_{\tau}^{\prime}|^2
+\frac{\Lambda_{\rm HC}^4}{\Lambda_N^4}|\lambda_N+\lambda_N^{\prime}|^2
\right] ,\label{eq:mH2L}\\
\Delta m_{12}^2=&\,\frac{a_f f^2}{64\pi^2}\bigg[
iN_c\frac{\Lambda_{\rm HC}^4}{\Lambda_t^4}(\lambda_t+\lambda_t^{\prime})(\lambda_t^{*}-\lambda_t^{\prime\,*})+
iN_c\frac{\Lambda_{\rm HC}^4}{\Lambda_b^4}(\lambda_b+\lambda_b^{\prime})(\lambda_b^{\prime\,*}-\lambda_b^{*})
\nonumber\\
&+\frac{i\Lambda_{\rm HC}^4}{\Lambda_N^4}(|\lambda_N^{\prime}|^2-|\lambda_N|^2)-\frac{4\Lambda_{\rm HC}^4}{\Lambda_{\tau}^4}{\rm Im}(\lambda_{\tau}^{*}\lambda_{\tau}^{\prime})
\bigg],\label{eq:m12L}
\end{align}
and contributions to the parameters of \cref{eq:VwithsH},
\begin{align}
\Delta \alpha_{+}=&-\frac{a_f }{16\pi^2 }\left(N_c \frac{\Lambda_{\rm HC}^4}{\Lambda_{t}^4} |\lambda_t|^2+
N_c \frac{\Lambda_{\rm HC}^4}{\Lambda_{b}^4} |\lambda^{\prime}_b|^2
+\frac{\Lambda_{\rm HC}^4}{\Lambda_{\tau}^4}( |\lambda_\tau|^2+|\lambda^{\prime}_\tau|^2)
- \frac{\Lambda_{\rm HC}^4}{\Lambda_{N}^4} |\lambda_N|^2
\right),\label{eq:alphap1}\\
\Delta \alpha_{-}=&-\frac{a_f }{16\pi^2 }\left(N_c \frac{\Lambda_{\rm HC}^4}{\Lambda_{t}^4}|\lambda^{\prime}_t|^2+
N_c \frac{\Lambda_{\rm HC}^4}{\Lambda_{b}^4} |\lambda_b|^2
+\frac{\Lambda_{\rm HC}^4}{\Lambda_{\tau}^4}( |\lambda_\tau|^2+|\lambda^{\prime}_\tau|^2)- \frac{\Lambda_{\rm HC}^4}{\Lambda_{N}^4} |\lambda^{\prime}_N|^2
\right),\label{eq:alpham1}\\
\Delta {\alpha_{\rm sin}}=&-\frac{a_f}{8\pi^2 }\left(N_c \frac{\Lambda_{\rm HC}^4}{\Lambda_{t}^4}{\rm Re}(\lambda_t^* \lambda_t^{\prime})
-N_c \frac{\Lambda_{\rm HC}^4}{\Lambda_{b}^4}{\rm Re}(\lambda_b^* \lambda_b^{\prime})+2\frac{\Lambda_{\rm HC}^4}{\Lambda_{\tau}^4}{\rm Re}(\lambda_{\tau}^* \lambda_{\tau}^{\prime})\right),\label{eq:alphas1}\\
\Delta {\alpha_{\rm cos}}=&-\frac{a_f }{8\pi^2 }\frac{\Lambda_{\rm HC}^4}{\Lambda_{N}^4} {\rm Re}(\lambda_N^* \lambda_N^{\prime}).\label{eq:alphac1}
\end{align}
The gauge symmetry restrict the Wilson coefficients $\lambda^{ (\prime)}$ of \cref{eq:EHCt} to 12 independent parameters: 6 real, $\lambda_{(LX)}$, 
$\lambda_{(3X)}$, $\lambda_{(4X)}$, and 6 complex, $\lambda^{\prime}_{(LX)}$, $\lambda^{\prime}_{(3X)}$, $\lambda^{\prime}_{(4X)}$, with $X=L,R$. They are arranged as
\begin{align}
 \lambda_{\rho_1\rho_2\rho_3\rho_4}= \lambda_{(XY)}\delta_{\rho_1 \rho_4} \delta_{\rho_2 \rho_3},~~~~
\lambda^{\prime}_{\rho_1\rho_2\rho_3\rho_4}= \lambda^{\prime}_{(XY)} \tilde\epsilon_{\rho_1\rho_3} \tilde\epsilon_{\rho_2 \rho_4},
\end{align}
where 
\begin{equation}
X=\begin{cases}
L~~\,{\rm if}~\rho_1=1,2 \\
\rho_1~~{\rm if}~\rho_1=3,4 
\end{cases},~~~~
Y=\begin{cases}
L~~\,{\rm if}~\rho_2=1,2 \\
R~~{\rm if}~\rho_2=3,4 
\end{cases},~~~~
\tilde \epsilon = \begin{pmatrix}
\epsilon_{2\times 2} && 0_{2\times 2} \\
0_{2\times 2} && \epsilon_{2\times 2}
\end{pmatrix},
\end{equation}
with $\epsilon_{2\times 2}$ the two-dimensional Levi-Civita tensor.
Expanding now \cref{eq:V0}, we additionally obtain the tadpole for $S$
\begin{equation}
V_{\rm scalar} \supset \frac{\sqrt{2}}{2}\frac{ \Lambda_{\rm HC}^4 }{\Lambda_{\zeta}^4}a^{\prime}f^3\, {\rm Im}(2\lambda_{(LL)}-\lambda_{(3R)}-\lambda_{(4R)}),\label{eq:STadp}
\end{equation}
and contributions to the pNGB masses
\begin{align}
\Delta m_S^2=&-\frac{a^{\prime}\Lambda_{\rm HC}^2f^2}{2\Lambda_{\zeta}^2}{\rm Re}\left( 2\lambda^{\prime}_{(LL)} +\lambda^{\prime}_{(3R)}+\lambda^{\prime}_{(4R)}
 \right), \label{eq:mST}\\
\Delta m_{H_1}^2=&\frac{\Lambda_{\rm HC}^2f^2}{4\Lambda_{\zeta}^2}\bigg[ a
\left(
2\lambda_{(LL)} - 2\lambda_{(LR)} - \lambda_{(3L)} -\lambda_{(4L)} 
+ \lambda_{(3R)} +\lambda_{(4R)}\right) 
\nonumber\\
&+2a^{\prime}
{\rm Re}\left(
2\lambda^{\prime}_{(LR)} + \lambda^{\prime}_{(3L)} +\lambda^{\prime}_{(4L)} 
-2\lambda^{\prime}_{(LL)} - \lambda^{\prime}_{(3R)} -\lambda^{\prime}_{(4R)} 
\right)\bigg]\label{eq:mH1T}
,\\
\Delta m_{H_2}^2=&\frac{\Lambda_{\rm HC}^2f^2}{4\Lambda_{\zeta}^2}
\bigg[ a\left(
2\lambda_{(LL)} -
2\lambda_{(LR)} - \lambda_{(3L)} -\lambda_{(4L)}  + \lambda_{(3R)} +\lambda_{(4R)} 
\right)\nonumber\\
&+2a^{\prime}
{\rm Re}\left(-2\lambda^{\prime}_{(LR)} 
 - \lambda^{\prime}_{(3L)} -\lambda^{\prime}_{(4L)} 
 -2\lambda^{\prime}_{(LL)} - \lambda^{\prime}_{(3R)} -\lambda^{\prime}_{(4R)}
\right)\bigg]
, \label{eq:mH2T}\\
\Delta m_{12}^2=&\frac{\Lambda_{\rm HC}^2f^2}{4\Lambda_{\zeta}^2}\bigg[
i\,a\left( 
\lambda_{(3L)}- \lambda_{(4L)} 
-\lambda_{(3R)}+\lambda_{(4R)}\right)
+2a^{\prime}{\rm Im}\left(2\lambda^{\prime}_{(LR)}-\lambda^{\prime}_{(3L)}-\lambda^{\prime}_{(4L)} \right)\bigg].\label{eq:m12T}
\end{align}
Notice that, if $\lambda_t=\pm\lambda_t^{\prime}$, $\lambda_b=\pm\lambda_b^{\prime}$, $\lambda_{\tau}=\pm \lambda_{\tau}^{\prime}$, $\lambda_{N}=\pm \lambda_{N}^{\prime}$, $\lambda_{(3L)}=\lambda_{(4L)}$, 
$\lambda_{(3R)}=\lambda_{(4R)}$ and $\lambda^{\prime}_{(XY)}\in \mathbb{R}$, then $\Delta m_{12}^2=0$.
These relations imply that the contributions to the pNGB potential in~\cref{eq:V1,eq:V0} are invariant under one of the transformations $U \to P U^{*} P^{\dagger}$, with $P \in SU(4)$,
\begin{equation}
P= \begin{pmatrix}
\epsilon_{2\times 2} && 0_{2\times 2} \\
0_{2\times 2} && \epsilon_{2\times 2}
\end{pmatrix}~~~{\rm or}~~~
P= \begin{pmatrix}
i\epsilon_{2\times 2} && 0_{2\times 2} \\
0_{2\times 2} &&-i \epsilon_{2\times 2}
\end{pmatrix}.
\end{equation}
In terms of the pNGBs, they become $H_1\to H_1$, $H_2\to -H_2$, $S\to -S$, or $H_1\to -H_1$, $H_2\to H_2$, $S\to -S$, respectively. 
Thus, any mass mixing between $H_1$ and $H_2$ indeed vanishes~\cite{Mrazek:2011iu}.
Finally, the parameters of \cref{eq:VwithsH} receive the four-hyper-quark contributions
\begin{align}
\Delta \alpha_{+}=&\frac{a \Lambda_{\rm HC}^2}{\Lambda_{\zeta}^2}\left(\lambda_{(LL)}+\lambda_{(3R)}-\lambda_{(LR)}-\lambda_{(3L)}\right),\label{eq:alphap0}\\
\Delta \alpha_{-}=&\frac{a \Lambda_{\rm HC}^2}{\Lambda_{\zeta}^2}\left(\lambda_{(LL)}+\lambda_{(4R)}-\lambda_{(LR)}-\lambda_{(4L)}\right),\label{eq:alpham0}\\
\Delta {\alpha_{\rm sin}}=&\frac{2a^{\prime} \Lambda_{\rm HC}^2}{\Lambda_{\zeta}^2}\, {\rm Re} \left(2 \lambda_{(LR)}^{\prime}+\lambda_{(3L)}^{\prime}+\lambda_{(4L)}^{\prime} \right),\label{eq:alphas0}\\
\Delta {\alpha_{\rm cos}}=&\frac{2a^{\prime} \Lambda_{\rm HC}^2}{\Lambda_{\zeta}^2}\, {\rm Re} \left(2 \lambda_{(LL)}^{\prime}+ \lambda_{(3R)}^{\prime}+\lambda_{(4R)}^{\prime}\right).\label{eq:alphac0}
\end{align}

\section{Some extended hyper-color sectors}
\label{sec:EHC}

Mediators necessary to generate the four-fermion operators that induce the third-family Yukawa couplings and the explicit breaking of the $SU(4)_1\times SU(4)_2$ symmetry for the scalar potential could be massive vectors or scalars~\cite{Ma:2015gra}. We here comment on some of the possibilities. We investigate adding three self-conjugated scalars in representations $\Phi_{LR}\sim({\bf 2},{\bf 1},{\bf 1},\bar{\bf 2})_0$, $\Phi_{RL}\sim({\bf 1},\bar{\bf 2},{\bf 2},{\bf 1})_0$ and $\Phi_{RR}\sim({\bf 1},{\bf 2},{\bf 1},\bar {\bf 2})_0$ under ${\cal G}_{2^41}=SU(2)_{L1}\times SU(2)_{R1}\times SU(2)_{L2}\times SU(2)_{R2}\times U(1)_{B-L}$, a (hyper)-colored scalar $\cal S\sim ({\bf N}_{\rm HC} ,\bar {\bf 3})$ under $SU(N_{\rm HC})\times SU(3)_c$ with $U(1)_X$-charge $1/2N_{\rm HC}-1/6$, and a hyper-colored vector, ${\cal U}_{\mu}$, in the fundamental of $SU(N_{\rm HC})$ and $U(1)_X$-charge $1/2N_{\rm HC}+1/2$. 
Note that our purpose here is not to add all of them, as in general they could break $U(2)_{u,d,\ell}$, as discussed below~\cref{eq:ModelI4ferYuk}, but to discuss the contributions of different possibilities. In particular, we should not add $\Phi_{RL}$ together with ${\cal S}$, or $\Phi_{LR}$ with ${\cal U}$.
Couplings consistent with the gauge symmetry are
\begin{align}
\mathcal{L}& \supset \bar q^3_L\,\Phi_{RL} \,(\,c_{t} t_R,\,c_{b} b_R)^t + c_{\tau} \, \bar \ell^3_L\,\Phi_{LR}\, \ell^3_R +c_{N} \,(\bar N_L,0)\,\Phi_{RR}   \ell^3_R \nonumber\\
&+
 (c_{3L}\, \bar \zeta^{(R)}_{L,1},\,c_{4L} \,\bar \zeta^{(R)}_{L,2})\,\Phi_{RL}^{\dagger} \,  \zeta_R^{(L)} +c_{LR} \, \bar \zeta_L^{(L)}\,\Phi_{LR} \, \zeta^{(R)}_R +( c_{3R} \, \bar \zeta_{L1}^{(R)}, c_{4R} \, \bar \zeta_{L,2}^{(R)})\,\Phi_{RR}\,   \zeta_R^{(R)}\nonumber\\
 &+( \bar \zeta^{(R)}_{L,1},\bar \zeta^{(R)}_{L,2}) \,{\cal S}\, (y_R^t t_R,y_R^b b_R)^t +  y_L \bar q^3_L {\cal S}^* \zeta_R^{(L)} \nonumber\\
 &+g_{\ell_L}\bar \ell^{\,3}_L\, \cancel{{\cal U}}^{}\zeta_{L}^{(L)}+g_{\ell_R}\bar \zeta_{R}^{(R)} \cancel{{\cal U}}^{}\ell^3_R+g_{N} \bar N_L\,\cancel{{\cal U}}^{}\zeta_{L,1}^{(R)} +{\rm h.c}.
\label{eq:EHCCouplings}
\end{align}
Assuming their masses are $M\gtrsim \Lambda_{\rm HC}$, we can integrate them out, obtaining contributions at the tree level to the four-fermion Wilson coefficients of \cref{eq:ModelI4ferYuk,eq:4FLambdanu3,eq:EHCt}:
\begin{align}
\frac{\lambda_{t}}{\Lambda_t^2}=&\frac{c_t c_{3L}}{M_{RL}^2}-\frac{y_L y_{R}^t}{2M_{\cal S}^2},~~\frac{\lambda^{\prime}_{t}}{\Lambda_t^2}=\frac{c_t c^*_{4L}}{M_{RL}^2},\\
\frac{\lambda_{b}}{\Lambda_{b}^2}=&\frac{c_b c_{4L}}{M_{RL}^2}-\frac{y_L y_{R}^b}{2M_{\cal S}^2},~~\frac{\lambda^{\prime}_{b}}{\Lambda_b^2}=-\frac{c_b c^*_{3L}}{M_{RL}^2},\\
\frac{\lambda_{\tau}}{\Lambda_{\tau}^2}=&\frac{c_{\tau} c_{LR}^*}{M_{LR}^2}+2\frac{g_{\ell_L} g_{\ell_R}}{M_{\cal U}^2},~~\frac{\lambda_{\tau}^{\prime}}{\Lambda_{\tau}^2}=\frac{c_{\tau} c_{LR}}{M_{LR}^2},\\
\frac{\lambda_{N}}{\Lambda_{N}^2}=&\frac{c_N c_{3R}^*}{M_{RR}^2}+2\frac{g_{N} g_{\ell_R}}{M_{\cal U}^2},~~\frac{\lambda^{\prime}_{N}}{\Lambda_{N}^2}=\frac{c_N c_{4R}}{M_{RR}^2},\\
\frac{\lambda_{(LR)}}{\Lambda_{\zeta}^2}=&\frac{|c_{LR}|^2}{M_{LR}^2},~~\frac{\lambda^{\prime}_{(LR)}}{\Lambda_{\zeta}^2}=\frac{c_{LR}^2}{2M_{LR}^2},\\
\frac{\lambda_{(iL)}}{\Lambda_{\zeta}^2}=&\frac{|c_{iL}|^2}{M_{RL}^2},~~\frac{\lambda^{\prime}_{(iL)}}{\Lambda_{\zeta}^2}=\frac{c_{iL}^2}{2M_{RL}^2},\\
\frac{\lambda_{(iR)}}{\Lambda_{\zeta}^2}=&\frac{|c_{iR}|^2}{M_{RR}^2},~~\frac{\lambda^{\prime}_{(iR)}}{\Lambda_{\zeta}^2}=\frac{c_{iR}^2}{2M_{RR}^2},\\
\lambda_{(LL)}=&\lambda^{\prime}_{(LL)}=0.
\end{align}
Notice that the contributions to $\Delta m_{12}^2$ from the scalar fields $\Phi$ vanish if the couplings satisfy $c^2_{LR},c^2_{iL},c^2_{iR}, \in {\mathbb R}$, $c_{3L}=\pm c^*_{4L}$ and $c_{3R}=\pm c^*_{4R}$.

If the hyper-colored fields ${\cal S}$ or ${\cal U}_{\mu}$ have a mass $M<\Lambda_{\rm HC}$, they should not be integrated out. Instead, they confine together with a hyper-quark forming a bound fermionic state that mixes linearly with a SM field through the couplings of the third and fourth lines of \cref{eq:EHCCouplings}, implementing partial compositeness~\cite{Sannino:2016sfx,Cacciapaglia:2017cdi,Sannino:2017utc,Agugliaro:2019wtf}. 
These couplings are spurions of the global group $SU(4)_1\times SU(4)_2$ transforming like $y_L\sim ({\bf 1},\bar {\bf 4}) $, $y^{t,b}_R\sim ({\bf 4}, {\bf 1})$,  $g_{\ell_L}\sim (\bar {\bf 4}, {\bf 1}) $, $g_{\ell_R}\sim ({\bf 1}, {\bf 4})$ and $g_{N}\sim  (\bar {\bf 4}, {\bf 1})$.
In this case, the Yukawa couplings and masses generated are the same than in~\cref{eq:Yuk,eq:YukawasH3,eq:Nu3Mass} of \cref{sec:Yukawa} with the substitutions
\begin{align}
\frac{\lambda_t}{\Lambda_{t}^2}\to \frac{y_Ly^t_R}{\Lambda_{\rm HC}^2},~\lambda^{\prime}_t\to0,\\
\frac{\lambda_b}{\Lambda_{b}^2}\to \frac{y_Ly^b_R}{\Lambda_{\rm HC}^2},~\lambda^{\prime}_b\to0,\\
\frac{\lambda_{\tau}}{\Lambda_{\tau}^2}\to \frac{g_{\ell_L}g_{\ell_R}}{\Lambda_{\rm HC}^2},~\lambda^{\prime}_{\tau}\to0,\\
\frac{\lambda_{N}}{\Lambda_{N}^2}\to \frac{g_{N}g_{\ell_R}}{\Lambda_{\rm HC}^2},~\lambda^{\prime}_{N}\to0.
\end{align}
Furthermore, their contributions to the scalar potential are the same than in \cref{eq:V1} of \cref{sec:Scalar} with the same substitutions and no extra terms appear.

In the case of the field ${\cal S}$, two insertions of $y_L$  or $y_R^t$ generate new operators in the chiral description:
\begin{align}
\mathcal{L} &\supsetsim \frac{ f^2 y_Ly_L^*}{{\rm max}(M_{\cal S}^2,\Lambda^2_{\rm HC})} {\rm Tr}(U^\dagger  i{D}_{\mu} U {\rm diag}(\sigma_a ,0 ) )\,  \bar q_L^3\sigma^a \gamma^{\mu} q^3_L \nonumber\\
&+ \frac{ f^2 y^t_Ry_R^{t\,*}}{{\rm max}(M_{\cal S}^2,\Lambda^2_{\rm HC})} {\rm Tr}(U^\dagger  {\rm diag}(0,\sigma_3 ) i{D}_{\mu} U)\,  \bar t_R \gamma^{\mu} t_R.
\end{align} 
which contribute to the operators of \cref{eq:HqOp}. 
The four-quark operators of \cref{eq:qqOp} could come from a loop of hyper-quarks and ${\cal S}$, or integrating out $\Phi_{RL}$, which generates ${\cal C}^{(1)}_{qt}$ at the tree level.

The six-fermion operators for light Yukawa couplings of \cref{eq:4ferLYuk} can be generated by adding hierarchically heavier 
new fields.
For instance, let us take a heavy  self-conjugated  scalar, $\Phi_1 \sim ({\bf 2},\bar {\bf 2},{\bf 1},{\bf 1})_{0}$ under ${\cal G}_{2^41}$, with mass $M_{\Phi_1}\gg \Lambda_{\rm HC}$ and couplings
\begin{equation}
\mathcal{L} \supset  c_{\Phi^3} M_{\Phi_1} \Phi_1 \Phi_{RR} \Phi_{LR}^{\dagger} + \sum_{\substack{i=1,2 \\j=1,2,3}}\bar q^{j}_L \Phi_{1} (y^u_{ij} u^j_R, y^d_{ij}d^j_R)^t +{\rm h.c}.
\end{equation}
When $\Phi_1$ is integrated out at the scale $M_{\Phi_1}$, it generates the effective operators
\begin{equation}
\mathcal{L} \supset  \frac{c_{\Phi^3}}{M_{\Phi_1}}
 \sum_{\substack{i=1,2 \\j=1,2,3}}  \bar q^i_L\,\Phi_{RL}  \Phi_{RR} \,  (y^u_{ij} u^j_R, y^d_{ij}d^j_R)^t +{\rm h.c}+\dots\label{eq:Proto6}
\end{equation}
These operators give some of the six-fermion operators that generate light-quark Yukawa couplings of \cref{eq:4ferLYuk}  when $\Phi_{RL}$ and $\Phi_{RR}$ are integrated out, with $\tilde \Lambda_{u,d}^5=M_{\Phi_1}M_{RL}^2M_{RR}^2$.
Heavy vector-like fermions at the scale $M_{\Phi_1}$ (for instance, $Q_{L,R} \sim ({\bf 1},{\bf 1},{\bf 1},{\bf 2})_{1/6}$) can also generate similar operators to \cref{eq:Proto6}.

\section{Matching to SMEFT}
\label{sec:ApSMEFT}

We integrate out the massive vector bosons at the tree level to dimension 6 in SMEFT and include the leading terms of the CCWZ Lagrangian of \cref{eq:H1DH2}, rotating with \cref{eq:ImRot} and reducing to the Warsaw basis~\cite{Grzadkowski:2010es, Gherardi:2020det},
\begin{equation}
\mathcal{L}\supset \sum_n { C}_n O_n.
\end{equation}
Working in the interaction basis and taking $i,j=1,2,3$ and $k=1,2$ for the flavor indices, we get at the matching scale the Wilson coefficients:
\begin{align}
C_{H\Box}=&\,-\frac{1}{8 f^2}-\frac{3\cos^2(2\theta_L)}{32f^2}-\frac{\cos^2(2\theta_X)}{32f^2},\\
C_{HD}=&\,\frac{\sin^2(\theta_X)\cos^2(\theta_X)}{2f^2}-\frac{c_H^2s_H^2}{2f^2},\\
[C^{(3)}_{Hl}]_{ii}=&\,\frac{\cos(2\theta_L)\sin^2(\theta_L)}{8f^2}, \\
[C^{(3)}_{Hq}]_{kk}=&\,\frac{\cos(2\theta_L)\sin^2(\theta_L)}{8f^2}, \\
[C^{(3)}_{Hq}]_{33}=&\,-\frac{\cos(2\theta_L)\cos^2(\theta_L)}{8f^2}, \\
[C^{(1)}_{Hq},C^{(1)}_{H\ell},C_{Hu},C_{Hd}]_{ii}=&\, \frac{Y_\psi}{4f^2} \cos(2\theta_X)\sin^2(\theta_X),~~~(\psi=q,\ell,u,d),\\
[C_{He}]_{kk}=&\,
-\frac{\cos(2\theta_X)\sin^2(\theta_X)}{4f^2}, \\
[C_{He}]_{33}=&\,
\frac{\cos^2(\theta_X)\cos(2\theta_X)}{8f^2},\\
[C_{\ell q}^{(3)}]_{iik k}=&\,-\frac{\sin^4 \theta_L}{4f^2}, \\
[C_{\ell q}^{(3)}]_{ii33}=&\,\frac{\sin^2 2\theta_L}{16f^2},\\
[C_{qq}^{(3)}]_{kkkk}=&\,-\frac{\sin^4 \theta_L}{8f^2}, \\
[C_{qq}^{(3)}]_{1122}=&\,-\frac{\sin^4 \theta_L}{4f^2}, \\
[C_{qq}^{(3)}]_{kk33}=&\,\frac{\sin^2 2 \theta_L}{16f^2}, \\
[C_{qq}^{(3)}]_{3333}=&\,-\frac{\cos^4 \theta_L}{8f^2},\\
[C_{\ell\ell}]_{iiii}=&\,-\frac{1}{8f^2}(\sin^4 \theta_L+\sin^4 \theta_X), \\
[C_{\ell \ell}]_{iijj}=&\,\frac{1}{4f^2}(\sin^4 \theta_L-\sin^4 \theta_X),~~~(i<j), \\
[C_{\ell\ell}]_{ijji}=&\,-\frac{\sin^4 \theta_L}{2f^2},~~~(i<j), \\
[C_{\ell q}^{(1)},C_{\ell u},C_{\ell d},C_{qu},C_{qd},C_{ud}]_{iijj}=&\,-\frac{Y_{\psi_1}Y_{\psi_2}\sin^4 \theta_X}{f^2},~~~(\psi_{1,2}=\ell,q,u,d), \\
[C_{qq}^{(1)},C_{uu},C_{dd}]_{iiii}=&\,-\frac{Y_{\psi}^2\sin^4 \theta_X}{2f^2},~~~(\psi=q,u,d), \\
[C_{qq}^{(1)},C_{uu},C_{dd}]_{iijj}=&\,-\frac{Y_{\psi}^2\sin^4 \theta_X}{f^2},~~~(i<j,~\psi=q,u,d),
\end{align}
\begin{align}
[C_{\ell e},C_{q e},C_{u e},C_{d e}]_{iikk}=&\,\frac{Y_{\psi}\sin^4 \theta_X}{f^2},~~~(\psi=\ell,q,u,d), \\
[C_{\ell e},C_{q e},C_{u e},C_{d e}]_{ii33}=&\,-\frac{Y_{\psi}\sin^2 2\theta_X}{8f^2},~~~(\psi=\ell,q,u,d),\\
[C_{ee}]_{kkkk}=&\,-\frac{\sin^4 \theta_X}{2f^2}, \\
[C_{ee}]_{1122}=&\,-\frac{\sin^4 \theta_X}{f^2}, \\
[C_{ee}]_{kk33}=&\,\frac{\sin^2 2\theta_X}{8f^2}, \\
[C_{ee}]_{3333}=&\,-\frac{\cos^4 \theta_X}{8f^2},\\
[C_{\psi H}]_{ij}=&\frac{(y_{\psi})_{ij}}{2}\left( -\frac{1}{3f^2} +\frac{\sin^2 (2 \theta_L)}{8f^2}+\frac{s_H^2 c_H^2}{6f^2}  \right),~(\psi=u,d,e),\\
C_{H}=& \lambda_H\left( -\frac{1}{3f^2} +\frac{\sin^2 (2 \theta_L)}{8f^2}+\frac{s_H^2 c_H^2}{6f^2} \right),
\end{align}
where $Y_{\psi}$ is the hypercharge of the field $\psi$, $(y_{\psi})_{ij}$ the SM Yukawa couplings and $\lambda_H$ the Higgs quartic, $\mathcal{L}\supset -\lambda_H |H|^4/2$.

\section{Oblique parameters}
\label{sec:AppOblique}

Under the assumption of universality defined in \cref{FN:Univ}, we can group the dimension 6 operators of the Warsaw basis with fermion currents into the universal operators:
\begin{align}
{\mathcal{L}}\supset \,\, C_{Hf}^{(1)} \,(H^{\dagger}i\overleftrightarrow{D}_{\mu}H) J_{B}^{\mu} 
+
C_{Hf}^{(3)} \,(H^{\dagger}\sigma_{a} i\overleftrightarrow{D}_{\mu}H)J_{W}^{a,\mu} 
+ C_{ff}^{(1)}\left( J_{B}^{\mu}  \right)^2
+ C_{ff}^{(3)}\left(J_{W}^{a,\mu}\right)^2,\label{eq:UnivOp}
\end{align}
where $J_{B,W}$ are the fermionic currents
\begin{align}
J_{B}^{\mu} =&\sum_{\substack{\psi=q,\ell,\\ u,d,e}}Y_{\psi} \sum_{i=1}^3 \bar \psi^i \gamma^{\mu} \psi^i ,\\
J_W^{a,\mu}=&\sum_{\psi=q,\ell}\sum_{i=1}^3 \bar \psi^i \frac{\sigma^a}{2} \gamma^{\mu} \psi^i,
\end{align}
and $\sigma_a$ are the Pauli matrices.
Expanding $J_{B,W}$ in \cref{eq:UnivOp} and Fierzing if necessary (for $[C_{\ell \ell}]$), one can easily read any Warsaw Wilson coefficient. For instance:
\begin{equation}
[C_{H\psi}^{(1)}]_{ij}=Y_{\psi} C_{Hf}^{(1)}\delta_{ij},~~[C_{H\psi}^{(3)}]_{ij}=\frac{1}{2} C_{Hf}^{(3)} \delta_{ij},~~[C_{\ell \ell}]_{1221}=C_{ff}^{(3)},~~[C_{\ell \ell}]_{1122}=\frac{1}{2}(C_{ff}^{(1)}-C_{ff}^{(3)}).
\end{equation}
The oblique parameters, as defined in Ref.~\cite{Barbieri:2004qk}, can be obtained going to the SILH basis, whose Wilson coefficients are easily relatable to the parameters~\cite{Giudice:2007fh}:\begin{align}
\hat S =& \frac{v^2}{2}\left(C_{Hf}^{(3)}+\frac{g_L^2}{g_Y^2} C_{Hf}^{(1)}-C_{ff}^{(3)}-\frac{g_L^2}{g_Y^2} C_{ff}^{(1)}+2g_L^2C_{HWB}\right),\\
\hat T=&-\frac{v^2}{2}\left( C_{HD}-2C_{Hf}^{(1)} +C_{ff}^{(1)}\right),\\
W =& -\frac{v^2}{2}C_{ff}^{(3)},\\
Y =& -\frac{v^2}{2}\frac{g_L^2}{g_Y^2}C_{ff}^{(1)}.
\end{align}
Notice that in the Warsaw basis, $\hat S$ and $\hat T$ not only receive contributions from the pure bosonic operators, but also from universal operators containing fermions.
As it can be seen in \cref{eq:WmassST,eq:deltaST}, the EWPOs at the $Z$ and $W$ poles only test the parameter $W$ and the combinations
\begin{align}
\hat S- Y =&\frac{v^2}{2}\left(C_{Hf}^{(3)}+\frac{g_L^2}{g_Y^2} C_{Hf}^{(1)}-C_{ff}^{(3)}+2g_L^2C_{HWB}\right),\\
\hat T-\frac{g_Y^2}{g_L^2}Y=&-\frac{v^2}{2}\left( C_{HD}-2C_{Hf}^{(1)} \right),
\end{align}
that can be mapped into the dimension 6 NP contributions to the parameters $\varepsilon_{1}$, $\varepsilon_{2}$ and $\varepsilon_{3}$ defined in Ref.~\cite{Barbieri:2004qk}. We also see that $C_{ff}^{(1)}$ does not play any role in the pole observables of the EW fit.

\bibliographystyle{JHEP}
\bibliography{references}

\end{document}